\documentclass[reprint,amsmath,amssymb,pra,superscriptaddress,english]{revtex4-2}
\usepackage{graphicx}
\usepackage{graphics}
\usepackage{dcolumn}
\usepackage{bm}
\usepackage{amsfonts}       
\usepackage{nicefrac}       
\usepackage{microtype}      
\usepackage{lipsum}
\usepackage{subcaption}
\usepackage{amsmath}
\usepackage{xcolor}
\usepackage{babel}
\def\ba{{\bf a}}
\def\bb{{\bf b}}

\def\b1{{1\!\!1}}

\def\cH{{\ca H}}

\def\cR{{\ca R}}

\def\bC{{\mathbb C}}           

\def\bR{{\mathbb R}}
\def\bP{{\mathbb P}}

\def\lae{{\lambda_{e}}}
\def\beq{\begin{eqnarray}}
\def\eeq{\end{eqnarray}}
\def\dt{{T_d }}

\newcommand{\ca}[1]{{\cal #1}}         

\newtheorem{theorem}{\em Theorem}[section]

\newtheorem{remark}[theorem]{\em Remark}

\DeclareMathOperator{\Tr}{Tr}

\begin{document}

\title{Entropy certification of a realistic QRNG based on single-particle entanglement}
\author{Sonia Mazzucchi \footnote{corresponding author: sonia.mazzucchi@unitn.it}}
\affiliation{%
 Department of Mathematics and TIFPA-INFN, University of Trento, Italy\\
}%

\author{Nicolò Leone}%
\affiliation{%
 Nanoscience Laboratory, Department of Physics, University of Trento, Italy\\
}%

\author{Stefano Azzini}%
\affiliation{%
 Nanoscience Laboratory, Department of Physics, University of Trento, Italy\\
}%

\author{Lorenzo Pavesi}%
\affiliation{%
 Nanoscience Laboratory, Department of Physics, University of Trento, Italy\\
}%

\author{Valter Moretti}
\affiliation{%
 Department of Mathematics and TIFPA-INFN, University of Trento, Italy\\
}%

\date{\today} 

\begin{abstract}
In single-particle entanglement (SPE) two degrees of freedom of a single particle are entangled. SPE  is a resource that can be exploited both in quantum communication protocols and in experimental tests of noncontextuality based on the Kochen-Specker theorem. SPE can be certified via a test of
quantum contextuality based on Bell inequalities.  Experiments of Bell-like inequality violation by single particle entangled systems may be affected by an analogue of the  locality loophole in this context, due to the presence of unavoidable non-idealities in the experimental devices which actually produce unwanted correlations between the two observables that are simultaneously measured. This issue is tackled here by quantitatively analyzing the behaviour of realistic devices in SPE experiments with photons. In particular, we show how it is possible to provide a semi-device independent randomness certification  of realistic quantum random number generators based on Bell inequality violation by SPE states of photons. The analysis is further enlarged to encompass, with a Markovian model, memory effects due to dead time, dark counts and afterpulsing affecting single photon detectors, in particular when not dealing with coincidence measurements. An unbiased estimator is also proposed for quantum transition probabilities out of the collection of experimental data.
\end{abstract}

\maketitle
\section{Introduction}

Entanglement is one of the fundamental features of quantum theory since it is able to produce correlations that do not have an analogue in classical physics. Initially pointed out as a  source of paradoxes \cite{EPR,Bell64}, it has recently gained a relevant role in the blooming ares of quantum information and quantum computing \cite{Ekert91,Bennett92,Ekert98,Macchiavello04}. From the mathematical point of view, the very definition of entanglement (or non-separability) relies upon the tensor product structure of the Hilbert space associated to the states of a quantum system and, in the simple case of a bipartite system, it is related to the notion of Schmidt  rank \cite{nielsen2002quantum}. From the physical point of view, we can distinguish between two kinds of entanglement. In {\em inter-particle entanglement} non-classical correlations are shared between the degrees of freedom of two different particles, while  {\em intra-particle} or {\em single-particle entanglement} (SPE)  involves independent degrees of freedom of a single particle such as momentum and polarization of a photon  \cite{Gadway09,valles2014generation} or momentum and spin of a massive particle \cite{hasegawa2004,hasegawa2011}. 


Analogously to interparticle entanglement,  even in the case of SPE the violation of Bell-type inequalities highlights  the presence of correlations between outcomes of  measurements that cannot be described by means of a realistic non-contextual hidden variable theory. To this regard, several experiments of Bell inequality violation by single-particle entangled states have been recently proposed \cite{simon2000,huang2003,hasegawa2011}.

Usually, 
 {\em Bell tests} are understood as  tests of {\em local} realism when {\em two} (or more) spatially separated parties of a quantum system are considered. Here, instead, there is a single particle, made of two independent subsystems, with an entangled state. In this case, the violation of the Bell inequality is  related with contextuality rather than non-locality.

In addition, SPE can be a resource in quantum information, as shown in \cite{saha2016robustness,Adhikari15,azzini2020single,Pasini20,azzini2020single}.
Advantages and disadvantages of inter-particle entanglement and SPE are discussed in \cite{Pasini20}, where the  experimental setup of fig. \ref{fig:Setup} is explicitly considered. We refer also to \cite{azzini2020single} for a review of the theory and few applications of SPE. Briefly, a clear advantage of SPE over inter-particle entanglement is the ease of production and the possibility of use of cheap sources such as an attenuated laser or a lamp. Moreover, in an experiment of Bell inequalities violation, the estimate of the quantum transition probabilities does not require coincidence detection. These facts allow to  raise significantly the efficiency of production of meaningful events contributing to the detection statistics. This feature and the robustness under environment-induced decoherence effects \cite{saha2016robustness} make SPE a resource in quantum information \cite{Adhikari15}. On the other hand, experiments of Bell inequality violation by SPE states necessarily rely on the fair sampling assumption, since the critical detection efficiency necessary to close the detection loophole \cite{larsson1998bell} is valid only in the case of inter-particle entanglement.

In addition,  tests of
quantum contextuality on SPE states are affected by issues arising from 
 non-idealities of the experimental devices that, in fact, produce unwanted correlations between the outcomes of measurements of observables related to independent degrees of freedom. Such effects are in principle absent in the case of Bell tests on interparticle entangled states due to the space-like separation of the two independent components, but they have to be  taken into account in the case of SPE states. 
 
 In this paper, we present a quantitative analysis of this issue in the case of optical experiments on violation of the  Clauser-Horne-Shimony-Holt (CHSH) inequality by SPE states of photons, in the lines of the experiment described in \cite{Pasini20}. 
 One of the main results of the paper is inequality \eqref{Ris-1}, which provides an upper analytical bound for  the threshold that the observed value of the  CHSH parameter $S $ (see Eq.\eqref{S-par}) has to exceed in order to actually certify entanglement, taking into account the non ideal behavior of the beam splitters that are employed in the experiment. 
 
This analysis plays a relevant role in the derivation of  a realistic bound for the device independent guessing probability in a quantum random number generator (QRNG)  based on CHSH inequality violation  via single particle entangled photons \cite{Leone21}.
As extensively discussed in \cite{Pironio10,Pironio13,acin2012randomness,acin2016certified}, entanglement and Bell-inequality violation provide a powerful tool for randomness certification in QRNG  based on a  Bell test. More precisely, the min-entropy $H_{\infty}$\cite{konig2009operational} of the raw data in a Bell experiment with measured CHSH parameter $S$ is  bounded from below by \cite{Pironio10}:
 \begin{equation}\label{Bound-Pironio}
     H_{\infty}\geq -\log_2\left(\frac{1}{2}+\frac{1}{2}\sqrt{2-\frac{S^2}{4}}\right)
 \end{equation}
The bound \eqref{Bound-Pironio} is {\em device-independent}, i.e., it does not depend on a particular modelling of the physical system realizing the QRNG. On the other hand, it requires a loophole-free Bell test which is not straightforward to realize in practical real-world implementations. As remarked in \cite{silman2013device}, the requirement of device-independence is too demanding and can be replaced by weaker assumptions adopted in modification of the original protocol presented in \cite{Pironio10}. To this end, Bell-inequality violation by single particle entanglement represents a  good trade-off between ease of experimental implementation and security. Indeed, as shown in the proof-of-principle experiment described in \cite{Leone21}, a QRNG based on SPE can be practically realized and a robust estimate of the corresponding min-entropy can be presented, relying upon a rather small set of assumptions on the features of the optical components. To this end, the second main result of the present paper is  inequality \eqref{bound-Hmin}, which modifies \eqref{Bound-Pironio} providing an analytical bound for the min-entropy produced by a QRNG based on SPE, such as the one reported in \cite{Leone21}.

The paper is organized as follows. In section \ref{sez2} a detailed  theoretical analysis of the realized experiment \cite{Pasini20} on  CHSH inequality violation for  SPE states of photons  is presented in the  case where all optical components do not deviate from the ideal behaviour. Section \ref{sez3} tackles the issues arising from realistic beam splitters and mirrors. In particular, it considers the case where  reflectance and transmittance depend explicitly on the polarization of the incoming photon,   causing unwanted correlations between the degrees of freedom of ``momentum" and ``polarization". In addition,  the presence of losses is taken into account. 
Section \ref{sezQRNG} deals with the application of the previous results to the entropy certification of a semi-device-independent QRNG based on Bell inequality violation.
In section \ref{sezMarkov} we present a Markovian model for the description of the memory effects due to the presence of detectors dead time and afterpulsing as well as a technique for the construction of an unbiased estimator for the quantum transition amplitudes. Finally, section VI concludes the paper.

\section{Bell inequality violation by SPE states} \label{sez2}

Let us  consider a quantum system  with two independent degrees of freedoms, denoted by A and B, each of them with two values, in such a way that the associated Hilbert space can be represented as the tensor product $\cH_A\otimes\cH_B$ with $\cH_A=\cH_B=\bC^2$. On each system  we perform the measurement of  observables $O^\ba:\cH_A\to \cH_A$ and $O^\bb:\cH_B\to \cH_B$  with two possible outcomes $x\in \{+1,-1\}$ and $y\in \{+1,-1\} $ respectively. In practice, such observables can be written in the form $O^\ba=\ba\cdot {\boldsymbol{\sigma}}=a_1\sigma_1+a_2\sigma_2+a_3\sigma_3$ and $O^\bb=\bb\cdot  \boldsymbol{\sigma}=b_1\sigma_1+b_2\sigma _2+b_3\sigma_3$ for suitable unit vectors $\ba, \bb\in \bR^3$ and $\sigma_i$, with $i=1,2,3$, are the Pauli matrices. We shall denote $P^\ba_{+1}, P^\ba_{-1}$ and $P^\bb_{+1}, P^\bb_{-1}$ the projectors belonging to the projection-valued measure (PVM) associated to $O^\ba=P^\ba_{+1}- P^\ba_{-1}$ and $O^\bb=P^\bb_{+1}- P^\bb_{-1}$, that can be represented as $P^{\ba}_{\pm 1}=\frac{1}{2}(I\pm \ba \cdot \boldsymbol{\sigma})$ and $P^{\bb}_{\pm 1}=\frac{1}{2}(I\pm \bb \cdot{\bf \boldsymbol{\sigma}})$. Denoting by $\rho$ a state of the compound system, we shall focus on the probabilities
 \begin{equation}\label{Q-dis}
 P(x,y|\rho,\ba,\bb):=\Tr[\rho  P^\ba_{x} \otimes P^\bb_{y}], \qquad x,y\in \{+1,-1\}.
 \end{equation}
 
The Bell inequality deals with two pairs of observables $O^\ba_i=\ba_i\cdot \boldsymbol{\sigma}$ and  $O^\bb_j=\bb_j\cdot \boldsymbol{\sigma}$, with $i,j=0,1$, on the subsystems $A$ and $B$ respectively.  
We henceforth assume that the four vectors $\ba_0,\ba_1, \bb_0,\bb_1$ are given.
If $\rho$  is  a quantum state on $\cH_A\otimes\cH_B$, we shall denote by  $E(\ba_i, \bb_j)$ the expected value of $O^\ba_i\otimes O^\bb_j$
\begin{multline}\label{corr-coef}
E(\ba_i, \bb_j)=\Tr[\rho O^\ba_i\otimes O^\bb_j]=P(+1, +1|\rho, \ba_i, \bb_j)\\+P(-1, -1|\rho, \ba_i, \bb_j) -P(-1, +1|\rho, \ba_i, \bb_j)\\ -P(+1, -1|\rho, \ba_i, \bb_j),
\end{multline}
while $S(\ba_0, \ba_1,\bb_0,\bb_1)$ will indicate the CHSH parameter
\begin{multline}\label{S-par}
S(\ba_0, \ba_1,\bb_0,\bb_1)=E(\ba_0, \bb_0)+E(\ba_0, \bb_1)+E(\ba_1, \bb_0)\\ -E(\ba_1, \bb_1).
\end{multline}
The CHSH inequality,
\begin{equation}
|S(\ba_0, \ba_1,\bb_0,\bb_1)|\leq 2\:,
\end{equation}
must hold whenever  the quantities $E(\ba_i, \bb_j)$ are interpreted as the expectation values of products of  random variables $X_i, Y_j$, $i,j=1,2$,
 defined on the same probability space. This physically corresponds to deal with a realistic non-contextual hidden variable theory where,  physically speaking, $X_i,Y_j$ play the role of {\em classical variables}.

\begin{figure}[h!]
\centering
\includegraphics{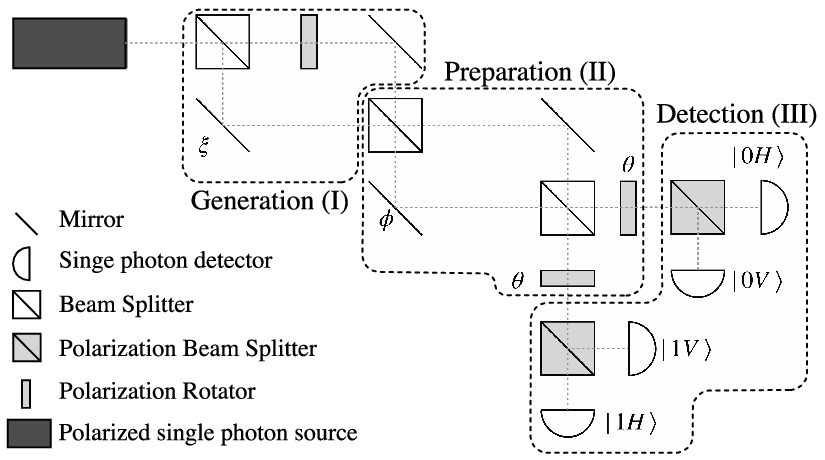}
\caption{Scheme of the setup used in \cite{Pasini20} to create momentum and polarization entangled photons and to perform the CHSH test of
quantum contextuality}. The grey dotted line represents the optical path of the photons. The black dotted lines show the three main parts of the setup: Generation (I), Preparation (II) and Detection (III). The angle $\xi$ is the angle used to correct phase mismatch in the SPE state, while the angles $\phi$ and $\theta$ are respectively, the rotation angles of the momentum and polarization states.
\label{fig:Setup}
\end{figure}

The  experimental implementation described in \cite{Pasini20} exploits momentum and polarization degrees of freedom of a single photon, which can be described as a 2-qubit system. We shall denote as $\{|0\rangle ,|1\rangle \}$ the vectors of an orthonormal basis in the momentum Hilbert space  $\cH_M$ corresponding to two particular propagation directions of the photon in the experimental setup. Analogously $\{|H\rangle ,|V\rangle \}$ denote the   vectors of the orthonormal basis in the polarization Hilbert space $\cH_P$ corresponding respectively  to horizontal and vertical polarization. Eventually, the corresponding orthonormal basis in the tensor product Hilbert space will be made of the four vectors $\{|0H\rangle,   |1H\rangle, |0V\rangle,|1V\rangle\}$, where henceforth $|XY\rangle =|X\rangle \otimes |Y\rangle$. From now on, unless otherwise stated, all the operators will be always represented by matrices in the above mentioned basis. As illustrated in \cite{Pasini20}, 
the experimental setup can be divided in three stages: (I)  generation, (II)  preparation,  and (III)  detection stage, as illustrated in Fig.\ref{fig:Setup}. 

The generation stage (I) has the role of transforming the state of a polarized single photon into the SPE entangled state $|\Psi_+\rangle=\frac{1}{\sqrt{2}}(|0H\rangle + i|1V\rangle)$. We consider the polarization of single photons entering the setup to be set to vertical, in such a way that the state vector of the photons can be represented as $|0V\rangle$. Next, a  beam splitter (BS), whose action on the vectors of $\cH_M$ can be represented by the matrix
\begin{equation}\label{VBS}
    V_{BS}=\left(\begin{array}{ll}
\sqrt{0.5}& i\sqrt{0.5} \\
i\sqrt{0.5}& \sqrt{0.5}
\end{array}\right),
\end{equation}
produces a superposition of momentum states of the form  $\frac{1}{\sqrt 2} (|0\rangle+i|1\rangle)\otimes |V\rangle$; finally  a half-wave plate rotates the polarization of photons with propagation direction $|0\rangle$, eventually producing  the SPE state  $|\Psi_+\rangle$.
The $i$ phase term is compensated by properly setting the phase $\xi$, which is controlled by varying the relative positions of the mirrors.
The preparation stage (II) consists of a Mach Zehnder interferometer (MZI) followed by two half-wave plates, one in each output port of the MZI. The MZI acts as a momentum-qubit gate $U_\ba:\cH_M\to \cH_M$. In the ideal case of a lossless balanced beam splitter, the unitary operator $U_\ba$ can be constructed out of the composition $U_\ba=V_{BS}
V_{mir}V(\phi)V_{BS}$, where $V_{BS}$ is given by \eqref{VBS} and 
\begin{equation}\label{Vphi} V(\phi)=\left(\begin{array}{ll}
e^{i\phi}&0 \\
0& 1
\end{array}\right),
\, V_{mir}=\left(\begin{array}{ll}
0&1 \\
1& 0
\end{array}\right).\end{equation}

\begin{remark}\label{REMMIR} The last operator
satisfies 
$$V_{mir}= V_{mir}^{-1} = V_{mir}^\dagger\:, $$
and represents the action of the mirrors between the two beam splitters (we assume they do not affect the polarization in any way). Notice that $V_{mir}$ commutes with $V_{BS}$ and that $V_{mir}\otimes I$  does the same with the analog operators
(\ref{BSM}),  which  describe  more realistic polarization-dependent and  lossy  beam splitters that we shall consider in the rest of the paper.
Therefore we have:
 $$U_\ba=V_{mir}V_{BS}V(\phi)V_{BS} \:.$$
 In addition, since $$V_{mir}P^M_{\pm}V_{mir}=P^M_{\mp}\:, $$
the net final effect of $V_{mir}$, 
 when composed with the projectors $P^M_{\pm}$,  is just to flip $P^M_{\pm}$ to $P^{M}_\mp$ 
without affecting the final results, since we did not  choose one of the two possible  one-to-one correspondence  between  the 
two possible labels $\pm 1$ and the momentum eigenstates $|0\rangle$ and $|1\rangle$. For this reason we shall henceforth omit $V_{mir}$ in the rest of computations and
assume
 $$U_\ba=V_{BS}V(\phi)V_{BS}\:.$$
 A more realistic mirror will be considered later in Section \ref{sez3}.
  \end{remark}

By explicit computation, we get
$$U_\ba= \left(\begin{array}{ll}
\frac{e^{i\phi}-1}{2}& \frac{i+ie^{i\phi}}{2} \\
\frac{i+ie^{i\phi}}{2} & \frac{1-e^{i\phi}}{2}
\end{array}\right)=ie^{i\phi/2}\left(\begin{array}{ll}
\sin(\phi/2)& \cos(\phi/2) \\
\cos(\phi/2)&- \sin(\phi/2)
\end{array}\right).$$
The last formula shows how the unitary operator $U_\ba$ associated to the MZI is  related to the phase shift $\phi $ in one arm of the Mach-Zehnder interferometer, where $ \phi = \frac{2\pi \Delta L}{\lambda}$,  $\Delta L$ is the path difference 
in the two arms and $\lambda$ the photon wavelength \cite{Gadway09}.
The angle $\phi$ determines the  vector $\textbf{a}$  in the Bloch sphere associated to  the 1-particle  observable $ O^{\bf a}={\bf a}\cdot \boldsymbol{\sigma}$ related to the momentum degree of freedom, where $\boldsymbol{\sigma} =(\sigma_x, \sigma_y, \sigma_z)$ are the associated Pauli matrices. In particular, the orthogonal projectors $\{P^\ba_{x}\}_{x=\pm1}$ associated to the eigenvectors of $O^{\bf a}$ are obtained by applying the unitary map $U_\ba$ to the projectors $\{P^M_{+1}=|0\rangle \langle 0|, P^M_{-1}=|1\rangle \langle 1|\}$  associated to the standard basis of $\mathcal{H}_M $, i.e. 
\begin{equation}\label{Pba} P^\ba_{x}=U_\ba ^\dag P^M_xU_\ba, \qquad x=\pm 1.\end{equation}

Next, two half-wave plates, one in each output port of the MZI (Fig. \ref{fig:Setup}), 
with the fast axis rotated by the same amount $\vartheta$, perform a rotation in the polarization space 
by an angle $\theta = 2\vartheta$ with respect to the vertical direction. This transformation in the qubit space $\mathcal{H}_P $ can be described in terms of a unitary map $U_\bb$. The  angle $\theta $ determines the vector
$\textbf{b}$  in the Bloch sphere, i.e. the 1-particle observable  $ O^{\bf b}={\bf b}\cdot{\bf \boldsymbol{\sigma}}$ related to the polarization degrees of freedom. As above, the orthogonal projectors $\{ P^\bb_{y} \}_{y=\pm 1}$ associated to the eigenvectors of $O^{\bf b}$ can be obtained as 
\begin{equation}\label{Pbb} P^\bb_{y}=U_\bb ^\dag P^P_y U_\bb, \qquad y=\pm 1,\end{equation} where $\{P^P_{+1}=|H\rangle \langle H|, P^P_{-1}=|1\rangle \langle 1|\}$ are the projectors associated to the basis $\{|H\rangle, |V\rangle\}$ of $\mathcal{H}_P $.
The net action of the preparation stage can be described by means of the unitary operator $U_{\ba, \bb}:=U_\ba\otimes U_\bb$ in the  space $\mathcal{H}_M \otimes \mathcal{H}_P$ transforming the state of an incoming photon, i.e. a density matrix $\rho$ in $\cH_M\otimes \cH_P$,  to the prepared state  $(U_\ba\otimes U_\bb)\rho (U_\ba\otimes U_\bb)^\dag$. 

The final detection stage (III) consists of two Polarizing Beam splitters (PBS) and four single-photon-avalanche-diodes (SPAD) that identify the four measurement channels and  correspond to the PVM made of the orthogonal projectors 
\begin{equation}|0H\rangle\langle 0H|, \quad  |0V\rangle\langle 0V|, \quad  |1H\rangle \langle 1H|, \quad  |1V\rangle\langle 1V|\label{PVM}\end{equation} 

Clearly, the projectors in \eqref{PVM} can be written as $P^M_x\otimes P^P_y$ where the superscrips M and P stands for momentum and polarizations,  while  $x, y\in \{+1, -1\}$. Actually $\{ P^M_{+1}, P^M_{-1}\}$ (resp. $\{ P^P_{+1}, P^P_{-1}\}$ ) are the PVM of the momentum observable $O^M=\sigma_z$ (resp. Polarization observable $O^P=\sigma_z$).

For any pair of unit vectors $\ba,\bb \in \bR^3$, the PVM associated to the joint measurement of the two commuting observables ${\bf a}\cdot{\bf \boldsymbol{\sigma}}\otimes I$ and $I\otimes {\bf b}\cdot{\bf \boldsymbol{\sigma}}$ is given by 
$$P_x^\ba\otimes P_y^\bb=U_\ba ^\dag P^M_x U_\ba \otimes U_\bb ^\dag P^P_yU_\bb =U_{\ba,\bb}^\dag(P_x^M\otimes P_y^P)U_{\ba,\bb}, $$ with  $x,y=\pm 1$.
 Given a state $\rho $ of the compound system, the corresponding quantum transition probabilities \eqref{Q-dis} are equal to \[P(x,y|\rho,\ba,\bb)=\Tr[\rho  P^\ba_{x} \otimes P^\bb_{y}]=\Tr[ U_{\ba,\bb}\rho U_{\ba,\bb}^\dag(P_x^M\otimes P_y^P)] \]

\section{Bell inequality violation by SPE states using realistic optical elements}\label{sez3}
\subsection{Issues with polarization-dependent transmittance and reflectance.  }
At this point it is worthwhile to stress that the correlations coefficients \eqref{corr-coef} and the CHSH parameter \eqref{S-par} refer to measurements of pair of commuting observables  of form ${\bf a}\cdot{\bf \boldsymbol{\sigma}}\otimes I$ and $I\otimes {\bf b}\cdot{\bf \boldsymbol{\sigma}}$  respectively. In the usual context of inter-particle entanglement this means that each observable is referred to a different particle of the entangled pair, while  in the case of SPE {\em each observable refers to a different degree of freedom of the same particle} (momentum or polarization in the specific experimental setting described above).  In particular, it is worth emphasising that this is a necessary condition for ruling out a description  of  CHSH  violation  in terms of a non-contextual hidden variables theory.  Furthermore,   protocols for device-independent entropy certification in quantum random number generators \cite{Pironio10} based on CHSH inequality violation rely upon the tensor product form $O^\ba={\bf a}\cdot{\bf \boldsymbol{\sigma}}\otimes I$ and $O^\bb=I\otimes {\bf b}\cdot{\bf \boldsymbol{\sigma}}$  of the couple of measured observables. 
In the case where the observables  $O^\ba$ and $O^\bb$ refer to space-like separated systems, this condition is naturally fulfilled, provided the Bell test is not affected by the locality loophole.  In the case of SPE, the situation is more complicated. In the  particular experimental implementation described above, the product form of the PVM $\{P^\ba_{x}\otimes P^\bb_{y}\}_{x,y=\pm 1}$ relies on the product form of the rotation operator  $U_\ba \otimes U_\bb$. As discussed above, this is obtained from the composition of $U_\ba \otimes I$,  realized by the Mach-Zehnder interferometer and acting only on the momentum degree of freedom, 
and of $I \otimes U_\bb$, realized by two half-wave plates and acting only on the polarization degree of freedom. While this last stage  does not present significant issues, in practical experimental implementations the beam splitters (BSs) employed in the MZI present a few non-idealities that  must be analyzed since they do not allow to represent the action of the interferometer in terms of a unitary operator of product form  $U_\ba \otimes I$. 
As we shall see shortly, this implies that the effectively measured observables  are not of the product form $A\otimes B$, giving rise to a problem analogous to the locality loophole  when dealing with couples of entangled particles.
In particular, in realistic BSs  reflectance and transmittance for    the vertically polarized component differ from those for the horizontally polarized  one. More specifically, the matrix representing the operator corresponding to the BS in the basis  $\{|0H\rangle,  |1H\rangle, |0V\rangle, |1V\rangle\}$ can be written as
\begin{equation}\label{BSM}
U_{BS}^{real}=\left(\begin{array}{llll}
t_H& ir_H & 0 &0\\
ir_H & t_H & 0 & 0\\
0 & 0 & t_V& ir_V\\
0 & 0 & ir_V& t_V
\end{array}\right)
\end{equation}
where $|t_H|^2+|r_H|^2\leq 1$ and $|t_V|^2+|r_V|^2\leq 1$. Actually, the operator $U_{BS}^{real}$ has the product form $V\otimes I$ if and only if 
\begin{equation}\label{idealBS}t_H=t_V, \quad \hbox{ and}\quad  r_H=r_V. \end{equation}  Hence, if  conditions \eqref{idealBS} are not fulfilled, the rotation operator  describing the action of the preparation stage $U^{real}_{\ba, \bb}=U_{BS}^{real}(V(\phi)\otimes I)U_{BS}^{real}$ ( with $V(\phi)$ defined in \eqref{Vphi}) can no longer be written as a tensor product $U_M\otimes U_{P}$, for a suitable pair of unitary operators $U_M:\cH_M\to \cH_M$ and $U_P:\cH_P\to \cH_P$. In other words, the probabilities of clicks of the four detectors in the final detection stage, namely
\begin{equation}\label{p-real} \Tr[U^{real}_{\ba,\bb}\rho (U^{real}_{\ba,\bb})^\dag \, P^M_{x}\otimes P^P_{y}]    \end{equation}
cannot be written in the form \begin{equation}\label{p-ideal}\Tr[\rho (P^{\tilde\ba}_x\otimes P_y^{\tilde\bb})]\end{equation} for a suitable pair of unit vectors $\tilde\ba, \tilde\bb \in \bR^3$  (also  different from $\ba,\bb$). To take this kind of non-idealities into account,  we provide a bound for the difference between the real probabilities \eqref{p-real} and the idealized ones \eqref{p-ideal} corresponding to tensor product observables. 
Furthermore, in order to handle the general case of a lossy beam splitter, we have also to consider  that a fraction of the photons passing through the Mach Zehnder interferometer performing the rotation in the momentum Hilbert space can actually be either scattered or absorbed. This means that the matrix \eqref{BSM} as well as the rotation operators $U^{real}_{\ba,\bb}$ constructed out of it  are not unitary   and the observed statistics of detection outcomes refers only to the photons that aren't lost by the optical elements. Hence, the actual  detection probabilities are computed as
\begin{equation}\label{p-real-2}
P^{real}(x,y|\rho,\ba,\bb)=\frac{ \Tr[U^{real}_{\ba,\bb}\rho (U^{real}_{\ba,\bb})^\dag \, P^M_{x}\otimes P^P_{y}]  }{\Tr[U^{real}_{\ba,\bb}\rho (U^{real}_{\ba,\bb})^\dag ] },
\end{equation}
where 
\begin{equation}\label{U-real-1}U^{real}_{\ba,\bb}=(I\otimes U_\bb)U_{BS}^{real}(V(\phi)\otimes I )U_{BS}^{real} .\end{equation}
If  the losses for the vertically polarized component are comparable  to the losses for the horizontally polarized one, i.e., when 
\begin{equation}\label{cond-semplif}t_H^2+r_H^2\sim t_V^2+r_V^2\end{equation}
then the denominator in \eqref{p-real-2} is equal to $t_H^2+r_H^2= t_V^2+r_V^2$  for any choice of $\rho, \ba, \bb$. This allows us to compute the detection probabilities \eqref{p-real-2} in terms  of the following expression
\begin{equation}\label{p-real-3}
P^{real}(x,y|\rho,\ba,\bb)=\Tr[\tilde U^{real}_{\ba,\bb}\rho (\tilde U^{real}_{\ba,\bb})^\dag \, P^M_{x}\otimes P^P_{y}] ,
\end{equation}with : 
\begin{equation}\label{utildereal}\tilde U^{real}_{\ba,\bb}= (I\otimes U_\bb)\tilde U_{BS}^{real}(V(\phi)\otimes I )\tilde U_{BS}^{real}  ,\end{equation}
where  the effective unitary operator $\tilde U_{BS}^{real}$ is defined as
$$   
\tilde U_{BS}^{real}=\left(\begin{array}{llll}
\tilde t_H& i\tilde r_H & 0 &0\\
i\tilde r_H & \tilde t_H & 0 & 0\\
0 & 0 & \tilde t_V& i\tilde r_V\\
0 & 0 & i\tilde r_V& \tilde t_V
\end{array}\right) ,$$    
\begin{eqnarray}\nonumber\tilde t_H&=&\frac{t_H}{\sqrt{t_H^2+r_H^2}} ,\quad   \tilde r_H=\frac{r_H}{\sqrt{t_H^2+r_H^2}},\\   \tilde t_V&=&\frac{t_V}{\sqrt{t_V^2+r_V^2}}, \quad \tilde r_V=\frac{r_V}{\sqrt{t_V^2+r_V^2}}.\label{normalization-0}\end{eqnarray}

In order to estimate the difference between the detection probabilities \eqref{p-real-3} and the ideal ones \eqref{p-ideal} associated to observables of product form, in the next two sub-sections we compute the unitary operator $U_\ba^{ideal}:\cH_M\otimes \cH_P\to\cH_M\otimes \cH_P$ of product form which minimizes the Hilbert-Schmidt distance from the operator $U_\ba^{real}:\cH_M\otimes \cH_P\to\cH_M\otimes \cH_P$  defined as $U_\ba^{real}=\tilde U_{BS}^{real}(V(\phi)\otimes I )\tilde U_{BS}^{real}$ when varying the parameters of the factors entering the expression of $U_\ba^{ideal}$ (see Eq. (\ref{Uid}) and Remark \ref{remaa} below).

\subsection{Equally lossy polarization channels}\label{secELPC}
We first consider the general case where the two BS included in the MZI  present different values of $t_H, t_V, r_H, r_V$.  Let us assume for the time being that  condition \eqref{cond-semplif}  is  still satisfied. This condition will be relaxed in the next sub-section. Under this assumption the operator $U_\ba^{real}$ can be written as 

\begin{equation}\label{MZ-general}
U_{\ba (\phi)}^{real}=V^{BS}_1(V(\phi)\otimes I)V^{BS}_2, 
\end{equation}
where \begin{equation}\label{V-BS-k}V^{BS}_k=\left(\begin{array}{llll}
t_{H,k}& ir_{H,k} & 0 &0\\
ir_{H,k} & t_{H,k}& 0 & 0\\
0 & 0 & t_{V,k}& ir_{V,k}\\
0 & 0 & ir_{V,k}& t_{V,k}
\end{array}\right), \; k=1,2.  \end{equation}
 By condition \eqref{cond-semplif} we can  restrict ourselves to the case where
\begin{equation}\label{normalization}t_{H,k}^2+r_{H,k}^2=t_{V,k}^2+r_{V,k}^2=1, \; k=1,2\end{equation}
since, if this condition is not fulfilled, the coefficients $t_H, r_H, t_V, r_v,$ can be replaced by the corresponding normalized coefficients $\tilde t_H, \tilde r_H, \tilde t_V, \tilde r_V$ as in \eqref{normalization-0}.
Hence, we can introduce the notation 
\begin{equation}
    \label{BS-2-real}V^{BS}_k=\left(\begin{array}{llll}
\cos \alpha^H_k& i\sin \alpha^H_k & 0 &0\\
i\sin \alpha^H_k & \cos \alpha^H_k& 0 & 0\\
0 & 0 & \cos \alpha^V_k& i\sin \alpha^V_k\\
0 & 0 & i\sin \alpha^V_k & \cos \alpha^V_k
\end{array}\right), \quad k=1,2. \end{equation}

We shall consider the difference between $U_{\ba(\phi)} ^{real}$ and a product operator $U_{\ba(\phi)} ^{ideal}$ of the form 
\begin{equation}
U_{\ba(\phi)} ^{ideal }(u,v)=(U(u)\otimes I)( V(\phi)\otimes I)(U(v)\otimes I) \label{Uid}
\end{equation}
where $u,v\in [0,2\pi]$ and 
\begin{equation}\label{V1xV_2y}U(\theta)=\left(\begin{array}{ll}
\cos \theta& i\sin \theta\\
i\sin \theta & \cos \theta
\end{array}\right), \end{equation}
and compute the values of $u,v$ that minimize the Hilbert-Schmidt norm of the difference operator $$R_{\ba}(u,v)=U_\ba ^{real}-U_\ba ^{ideal}(u,v)$$ adopting the shortened notation $\ba \equiv \ba (\phi)$. 

\begin{remark}\label{remaa}
The operator $U^{ideal}_{\ba}(u,v)$ corresponds to an initial  setup where the two BS in the preparation stage of Fig. \ref{fig:Setup} are ideal, generally non-balanced,  and different to each other  if $u\neq v$. The use of this unitary operator is equivalent  to a final measurement of the momentum observable $$\ba'(\phi,u,v) \cdot \boldsymbol{\sigma} = (U^{ideal}_{\ba(\phi)}(u,v))^\dagger\sigma_z U^{ideal}_{\ba(\phi)}(u,v) $$
for a unit vector $\ba'=\ba'(\phi,u,v)$ fulfilling the identity above, instead of the initially chosen  observable $\ba(\phi) \cdot \boldsymbol{\sigma}$. In other words, when taking the said  non-idealities into account but assuming (\ref{cond-semplif}), our setup is viewed to measure 
the observable 
\begin{equation}\ba'(\phi,u_0,v_0)\cdot \boldsymbol{\sigma} \otimes \bb\cdot \boldsymbol{\sigma}\:,\label{eqremaa}\end{equation}
where $(u_0,v_0)$ minimizes $||R_{\ba}(u,v)||_{HS}$,
instead of measuring  $\ba(\phi)\cdot \boldsymbol{\sigma} \otimes \bb \cdot \boldsymbol{\sigma}$. 
\end{remark}

 The operator $U_{\ba (\phi)}^{real}$ can be written in the block form 
\begin{equation}U_{\ba (\phi)}^{real}=e^{i\phi/2} \left(\begin{array}{ll}
U(\theta_H, \hat n_H) & 0 \\
0 & U(\theta_V, \hat n_V)
\end{array}\right) \label{U-real-block}\end{equation}
where 
\begin{align*}
U(\theta_H, \hat n_H) &=e^{i\theta_H \, \hat n_H \cdot \boldsymbol{\sigma}}=\cos \theta\,  I_{2\times 2}+i\sin \theta_H \, \hat n_H \cdot \boldsymbol{\sigma}\\
U(\theta_V, \hat n_V) &=e^{i\theta_V \, \hat n_V \cdot \boldsymbol{\sigma}}=\cos \theta_V\,  I_{2\times 2}+i\sin \theta_V \, \hat n_V \cdot \boldsymbol{\sigma}
\end{align*}
\begin{align*}
\cos\theta_H &= \cos \phi/2 \cos(\alpha^H_1 +\alpha^H_2),\\
  \sin\theta _H\hat n_H&=(\cos \phi/2 \sin(\alpha^H_1 +\alpha^H_2), \\
&\sin \phi/2 \sin(\alpha^H_1 -\alpha^H_2),\sin \phi/2 \cos(\alpha^H_1 -\alpha^H_2))\\
\cos\theta_V &= \cos \phi/2 \cos(\alpha^V_1 +\alpha^V_2),  \\
\sin\theta _V\hat n_V&=(\cos \phi/2 \sin(\alpha^V_1 +\alpha^V_2),\\ 
&\sin \phi/2 \sin(\alpha^V_1 -\alpha^V_2),\sin \phi/2 \cos(\alpha^V_1 -\alpha^V_2)).
\end{align*}
Analogously,  $U_\ba ^{ideal}(u,v)=e^{i\phi/2}e^{i\tilde\theta(u,v) \, \hat n(u,v) \cdot \boldsymbol{\sigma}}\otimes I$, with 
\begin{multline}\cos\tilde\theta(u,v)= \cos \phi/2 \cos(u+v),\\ \sin\tilde\theta(u,v)\hat \tilde n(u,v)=(\cos \phi/2 \sin(u,v), \sin \phi/2 \sin(u-v),\\
\sin \phi/2 \cos(u-v)).
\end{multline}
Hence \begin{multline}\| R_\ba\|^2_{HS}=\Tr[R_\ba R^\dag_\ba]=4\big(2-\cos^2 (\phi/2)(\cos(\alpha_1^H+\alpha_2^H -u-v)\\+\cos(\alpha_1^V+\alpha_2^V -u-v))\\  -\sin^2 (\phi/2)(\cos(\alpha_1^H-\alpha_2^H -u+v)+\cos(\alpha_1^V-\alpha_2^V -u+v))\big)\end{multline}
and this quantity attains its minimal value for $u=\frac{\alpha_1^H +\alpha _1^V}{2}$, and $ v=\frac{\alpha_2^H +\alpha_2^V}{2}$. In particular, in this case the operator $R_\ba R^\dag_\ba$ is a multiple of the identity, i.e. $R_\ba R^\dag_\ba=cI_{4\times 4}$ with 
\begin{multline}
    c=2-2\cos^2 (\phi/2)\cos\left(\frac{\alpha_1^H-\alpha _1^V}{2}+\frac{\alpha_2^H- \alpha _2^V}{2}\right)\\ -2\sin^2 (\phi/2)\cos\left(\frac{\alpha_1^H-\alpha _1^V}{2}-\frac{\alpha _2^H- \alpha_2^V}{2}\right)
\end{multline}
Hence $\|R_\ba\|=\sqrt c$.
A  uniform bound on  the norm of $R_\ba R^+_\ba$ that is independent of $\ba$, hence of $\phi$, is given by
\begin{multline}\label{error-1}
e:=\sup_{\phi\in [0, 2\pi]}\|R_\ba(\phi)R_\ba(\phi)^\dag\|\\
=2-2\min\Big\{\cos\left(\frac{\alpha_1^V-\alpha_1^H }{2}+\frac{\alpha_2^V- \alpha_2^H}{2}\right),\\ \cos\left(\frac{\alpha_1^V-\alpha_1^H }{2}-\frac{\alpha_2^V- \alpha_2^H}{2}\right)  \Big\}\end{multline}
\begin{remark}
The optimization technique implemented in this section actually allows to obtain an even  sharper bound. Indeed, it provides a value that coincides with the one attainable by the more general technique based on the comparison between  the rotation operator $U_{\ba(\phi)} ^{real}=V^{BS}_1  (V(\phi)\otimes I) V^{BS}_2$ associated to the MZ with a general unitary operator of product form $U_M\otimes V_P$, with $U_M, V_P$ unitary operators on $\cH_M$ and $\cH_P$ respectively.  More precisely, if $V^{ideal}_1 (x)$ and $V^{ideal}_2(y)$ are the unitary operators defined by \eqref{V1xV_2y}, we have:
\begin{multline}\label{equalitymaxmin}
\max_{\phi\in [0, 2\pi]}\min_{u,v\in [0,2\pi]}\|U_{\ba(\phi)} ^{real }-V^{ideal}_1 (u) (V(\phi)\otimes I) V^{ideal}_2(v)\|_2\\= \max_{\phi\in [0, 2\pi]}\min_{\zeta\in [0,2\pi],U_M,U_P \in SU(2)}\|U_{\ba(\phi)} ^{real }-e^{i\zeta}U_M\otimes U_P\|_2\, .
\end{multline}
A detailed proof of this result is given in appendix.
\end{remark}

We can now compute a uniform bound (independent of $\ba, \bb$) for the difference between the detection probabilities \eqref{p-real-2} and the ideal ones \eqref{p-ideal} associated to tensor product observables.

Indeed, by  expanding the expression $U^{real}_{\ba,\bb}=(I\otimes U_\bb) (U_\ba^{ideal}+R) $ we obtain:
\begin{multline}
P^{real}(x,y|\rho,\ba,\bb)
=P^{ideal}(x,y|\rho,\ba,\bb)\\ +\Tr[U_\ba^{ideal}\rho R_\ba^\dagger P^M_{x}\otimes P^P_{y,\bb}]+\Tr[R_\ba\rho (U_\ba^{ideal})^\dagger P^M_{x}\otimes P^P_{y,\bb}] \\+\Tr[R_\ba\rho R_\ba^\dagger P^M_{x}\otimes P^P_{y,\bb}]
\end{multline}
where $P^M_{x}\otimes P^P_{y,\bb}= (I\otimes U_\bb)^\dagger P^M_{x}\otimes P^P_{y} (I\otimes U_\bb)$ and  $P^{ideal}(x,y|\rho,\ba,\bb)=Tr[U_\ba^{ideal}\rho U_\ba^{ideal})^\dagger P^M_{x}\otimes P^P_{y, \bb}] $.

\begin{remark}
In spite of the notation $P^{ideal}(x,y|\rho,\ba,\bb)$ 
used here and in the next section, this
is the probability to get the outcomes $x$ and $y$ when measuring the factorized observable 
$$\ba'(\phi,u_0,v_0)\cdot \boldsymbol{\sigma} \otimes \bb\cdot \boldsymbol{\sigma}\:,$$
and not $\ba(\phi)\cdot \boldsymbol{\sigma} \otimes \bb\cdot \boldsymbol{\sigma}$,
according to Remark \ref{remaa}.
\end{remark}

For any choice of $\ba, \bb$,  the difference between the detection probabilities $P^{real}(x,y|\rho,\ba,\bb)$ and the ones related to product observables $ P^{ideal}(x,y|\rho,\ba,\bb)$ can be bounded by
\begin{multline}
    |P^{real}(x,y|\rho,\ba,\bb)-P^{ideal}(x,y|\rho,\ba,\bb)|\\
    \leq 2\sqrt{\|R_\ba R_\ba^\dagger\|}+\|R_\ba R_\ba^\dagger\|
    \leq 2 \sqrt e +e,
\end{multline}
where $e$ is given by \eqref{error-1}.

A similar bound can be derived for the difference between the CHSH parameter \eqref{S-par} associated to the detection probabilities \eqref{p-real-2} 
$$S^{real}=\Tr[\rho\sum_{a, b}c_{ab}(U_{\ba}^{real})^\dagger(I\otimes U_\bb^\dagger)\sigma_3\otimes \sigma_3(I\otimes U_\bb)U_{\ba}^{real}] $$
and an ideal one associated to measurements of product observables on the same state $\rho$
$$S^{ideal}= \Tr[\rho\sum_{\ba, \bb}c_{\ba\bb}(U_{\ba}^{ideal})^\dagger(I\otimes U_\bb^\dagger)\sigma_3\otimes \sigma_3(I\otimes U_\bb)U_{\ba}^{ideal}] ,$$
Where $c_{\ba_0,\bb_0}=c_{\ba_1,\bb_0}=c_{\ba_0,\bb_1}=1$ and $c_{\ba_1,\bb_1}=-1$.

As discussed above, for any choice of the vector $\ba$ associated to the phase shift $\phi$
 the difference operator $R_\ba=U_\ba ^{real}-U_\ba ^{ideal}$ can be written as 
\begin{equation}
\label{Ragen}
R_\ba=e^{i\phi/2}\left(f_1(\phi)R_1+f_2(\phi)R_2\right)
\end{equation}
where $f_1(\phi)=\cos(\phi/2)$, $f_2(\phi)=\sin(\phi/2)$ and the two operators $R_1,R_2$ do not depend on $\phi$ and are given by

$$R_1=\left(\begin{array}{ll}
R^H_1 & 0 \\
0 & R^V_1
\end{array}\right), \quad R_2=\left(\begin{array}{ll}
R^H_2 & 0 \\
0 & R^V_2
\end{array}\right)$$
where 
\begin{align*}
R^H_1 &=(\cos(\alpha_1^H +\alpha_2^H) -\cos(u+v))I_{2\times 2}+i(\sin(\alpha_1^H +\alpha_2^H)\\
& \qquad \qquad \qquad \qquad \qquad \qquad  \qquad \qquad  -\sin(u+v))\sigma_x\\
R^H_2 &=i(\cos(\alpha_1^H -\alpha_2^H) -\cos(x -y ))\sigma_z+i(\sin(\alpha_1^H -\alpha_2^H) \\ 
& \qquad \qquad \qquad \qquad \qquad \qquad  \qquad \qquad -\sin(u-v ))\sigma_y\\
R^V_1 &=(\cos(\alpha_1^V +\alpha_2^V) -\cos(u+v))I_{2\times 2}+i(\sin(\alpha_1^V +\alpha_2^V)\\
& \qquad \qquad \qquad \qquad \qquad \qquad  \qquad \qquad -\sin(u+v))\sigma_x\\
R^V_2 &=i(\cos(\alpha_1^V -\alpha_2^V) -\cos(u-v))\sigma_z+i(\sin(\alpha_1^V -\alpha_2^V) \\ 
& \qquad \qquad \qquad \qquad \qquad \qquad  \qquad \qquad -\sin(u-v))\sigma_y
\end{align*}
with  $u=(\alpha_1^H +\alpha_1^V)/2$ and $v=(\alpha_2^H +\alpha_2^V)/2$.  In the case the two beam splitters have similar features, i.e. $\alpha_1^H \sim \alpha_2^H$ and $\alpha _1^V\sim \beta _2^V$ then $R_2=0$. Since $R_1R_2^\dag+R_1^\dag R_2=0$ we have $\|R_\ba\|^2_{HS}=f_1^2(\phi)\|R_1\|^2_{HS}+f_2^2(\phi)\|R_2\|^2_{HS}$. Moreover 
\begin{align*}R_1R_1^\dag&=R_1^\dag R_1=4 \sin^2\left(\frac{\alpha_1^H +\alpha_2^H-\alpha_1^V-\alpha_2^V}{4}\right)I_{4\times 4}, \\
 R_2R_2^\dag&=R_2^\dag R_2=4 \sin^2\left(\frac{(\alpha_1^H -\alpha_2^H)-(\alpha_1^V-\alpha_2^V)}{4}\right)I_{4\times 4}\, ,
 \end{align*}
 hence
  \begin{align}
 \|R_1\|&=2\left|\sin\left(\frac{\alpha _1^H+\alpha_2^H-\alpha_1^V-\alpha_2^V}{4}\right) \right|\nonumber\\
  \|R_2\|&=2\left|\sin\left(\frac{(\alpha_1^H -\alpha_2^H)-(\alpha_1^V-\alpha_2^V)}{4}\right) \right|\label{R1R2norm}
 \end{align}
 
By expanding $U_{\ba}^{real}=U_{\ba}^{ideal}+R_\ba$ we get  
\begin{multline*}
S^{real}=S^{ideal}+\Tr[\rho\sum_{a, b}c_{ab}(R_\ba )^\dagger\sigma_3\otimes \bb\cdot \boldsymbol{\sigma} U_{\ba}^{ideal}] \\+\Tr[\rho\sum_{a, b}c_{ab}(U_{\ba}^{ideal})^\dagger\sigma_3\otimes \bb \cdot \boldsymbol{\sigma} R_\ba ] 
\\ +\Tr[\rho\sum_{a, b}c_{ab}R_\ba^\dagger\sigma_3\otimes \bb\cdot \boldsymbol{\sigma} R_\ba] 
\end{multline*}
Where $R_\ba$ is given by \eqref{Ragen}. In particular:

\begin{multline*} 
\Tr[\rho\sum_{a, b}c_{ab}(R_\ba )^\dagger\sigma_3\otimes \bb\cdot \boldsymbol{\sigma} U_{\ba}^{ideal}] \\ =\Tr[\rho R_1^\dag\sum_{a, b}c_{ab}e^{-i\phi/2}f_1(\phi(\ba)) \sigma_3\otimes \bb\cdot \boldsymbol{\sigma} U_{\ba}^{ideal}] \\+\Tr[\rho R_2^\dag\sum_{a, b}c_{ab}e^{-i\phi/2}f_2(\phi(\ba)) \sigma_3\otimes \bb\cdot \boldsymbol{\sigma} U_{\ba}^{ideal}] 
\end{multline*}
Each term on the right hand side has  the form $\Tr[\rho R_i^\dag O_i]$, with $i=1,2$ and \begin{multline*}  O_i=\sum_{a, b}c_{ab}e^{-i\phi/2}f_i(\phi(\ba)) \sigma_3\otimes \bb\cdot \boldsymbol{\sigma} U_{\ba}^{ideal}\\ 
=A_0B_0+A_0B_1+A_1B_0-A_1B_1\end{multline*}
with $B_j=I\otimes \bb_j\cdot \boldsymbol{\sigma}$ and  $A_j=e^{-i\phi/2}f_i(\phi(\ba_j)) (\boldsymbol{\sigma} \otimes I) U_{\ba}^{ideal}$, $j=0, 1$.
By explicit computation we have:
\begin{align*}
    \|O_iO_i^\dagger\|&\leq 2c_0^2+2c_1^2+2 |c_0^2-c_1^2|+4|c_0c_1|\\
    &\leq  \max_{|c_0|,|c_1| \in [0,1]}(2c_0^2+2c_1^2+2|c_0^2-c_1^2|+4c_0c_1)\\ &=8
\end{align*}
with $c_j:=f_i(\phi(\ba _j))$. 
Hence, by Von Neumann's trace inequality
\begin{equation*}
   |\Tr[\rho R_i^\dagger O_i]|\leq  \|R_i^\dagger O\|\Tr[\rho]\leq 2\sqrt 2\sqrt{\| R_i^\dagger R_i\|}.
\end{equation*}
and \begin{multline*}
    \Tr[\rho\sum_{a, b}c_{ab}(R_\ba )^\dagger\sigma_3\otimes \bb\cdot \boldsymbol{\sigma} U_{\ba}^{ideal}]\\ \leq 2\sqrt 2(\sqrt{\| R_1^\dagger R_1\|}+\sqrt{\| R_2^\dagger R_2\|}).
\end{multline*}
The same bound holds for the term $\Tr[\rho\sum_{a, b}c_{ab}(U_{\ba}^{ideal})^\dagger\sigma_3\otimes \bb \cdot \boldsymbol{\sigma} R_\ba ] $.
Similarly 
\begin{equation}\Tr[\rho\sum_{a, b}c_{ab}R_\ba^\dagger\sigma_3\otimes \bb\cdot \boldsymbol{\sigma} R_\ba] =\sum_{i,j=1,2}\Tr[R_j\rho R_i^\dag O_{ij}]\label{S-stima-3}\end{equation}
with $$O_{ij}=\sum_{a, b}c_{ab}f_i(\phi(\ba))f_j(\phi(\ba) \sigma_3\otimes \bb \cdot \boldsymbol{\sigma}.$$
Since $\|O_{ij}O_{ij}^\dagger\|\leq 2c_0^2+2c_1^2+2 |c_0^2-c_1^2|$, with $c_k=f_i(\phi(\ba_k))f_j(\phi(\ba_k)$, $k=0,1$, we have $\|O_{ij}\|\leq 2$ if $i=j$ and $\|O_{ij}\|\leq 1 $ if $i\neq j$, hence the right hand of \eqref{S-stima-3} is bounded by $2(\|R_1R_1^\dagger\|+\|R_2R_2^\dagger\|+\|R_1\|\|R_2\|)$. Eventually we have:
\begin{multline}\label{final-S-estimate-1}
    |S^{real}-S^{ideal}|\leq 4\sqrt 2(\|R_1\|+\|R_2\| ) \\ +2(\|R_1\|^2+\|R_2\|^2+\|R_1\|\|R_2\|)
\end{multline}
where $\|R_1\|$ and $\|R_2\|$ are given by \eqref{R1R2norm}.
\subsection{Generally lossy beam splitters}\label{sez4}

Let us consider now the general case where the two BS in the MZI providing the rotation of momentum qubit present different values of $t_H, t_V, r_H, r_V$ and  condition \eqref{cond-semplif}  is not satisfied. Hence, the approximations adopted in the previous section, in particular \eqref{p-real-3} and \eqref{utildereal},  are no longer feasible.
The detection probabilities are still given by  \eqref{p-real-2}, with $U^{real}_{\ba,\bb}=(I\otimes \bb\cdot \boldsymbol{\sigma})U^{real} _{\ba (\phi)}$, with $U^{real} _{\ba (\phi)}$ described by \eqref{MZ-general} and \eqref{V-BS-k}, but in the realistic case both BS present losses, which depend explicitly on the polarization:
\begin{equation}\label{losses}t_{H,k}^2+r^2_{H,k}\neq t_{V,k}^2+r^2_{V,k}, \quad k=1,2.\end{equation}
In fact, condition \eqref{losses} does not allow to get rid of the denominator in \eqref{p-real-2} and obtain  formula   \eqref{p-real-3}.
In the following we are going to estimate an upper bound for the difference between the detection probabilities \eqref{p-real-2} and the simplified ones
\begin{equation}
\Tr[\tilde U^{real}_{\ba,\bb}\rho (\tilde U^{real}_{\ba,\bb})^\dag \, P^M_{x}\otimes P^P_{y}] ,
\end{equation}with : 
\begin{equation}\label{utildereal-bis}\tilde U^{real}_{\ba,\bb}= (I\otimes U_\bb)\tilde U^{BS}_{1}(V(\phi)\otimes I )\tilde U^{BS}_{2}  .\end{equation}
where  the unitary operators $\tilde U^{BS}_{1}$ and $\tilde U^{BS}_{2}$ are defined as
\begin{equation} \label{UBStilde}
\tilde U^{BS}_{k}=\left(\begin{array}{llll}
\tilde t_{H,k}& i\tilde r_{H,k}& 0 &0\\
i\tilde r_{H,k} & \tilde t_{H,k} & 0 & 0\\
0 & 0 & \tilde t_{V,k}& i\tilde r_{V,k}\\
0 & 0 & i\tilde r_{V,k}& \tilde t_{V,k}
\end{array}\right) ,
\end{equation} where  $\tilde t_{H,k}=\frac{t_{H,k}}{\sqrt{t_{H,k}^2+r_{H,k}^2}}$ ,   $\tilde r_{H,k}=\frac{r_{H,k}}{\sqrt{t_{H,k}^2+r_{H,k}^2}}$,   $\tilde t_{V,k}=\frac{t_{V,k}}{\sqrt{t_{V,k}^2+r_{V,k}^2}}$, $\tilde r_{V,k}=\frac{r_{V,k}}{\sqrt{t_{V,k}^2+r_{V,k}^2}}$, $k=1,2$.

\begin{remark}\label{REMMIR2}
The above formula may include the contribution of the pair of  realistic mirrors in the preparation stage of Figure \ref{fig:Setup}. Each mirror is here permitted to be
lossy (with
losses depending on the polarization channel) but we assume that the two mirrors have very similar physical characteristics. The difference with the ideal case discussed in Remark \ref{REMMIR}, is just that the matrix representing the couple of  mirrors has a further numerical factor $\eta\in (0,1)$ in front of the  unitary matrix $V_{mir}$ in (\ref{Vphi}). When passing to the description in terms of $4\times 4$ complex matrices as in (\ref{BSM}) to take the two polarization into account, the factor may be different for the two $2\times 2$ matrices on the principal diagonal and  we may have two coefficients $\eta_H, \eta_V \in (0,1)$.  The net final  effect of these two further factors is just to rescale the coefficients
appearing in (\ref{V-BS-k}) with $k=2$ to
\begin{equation}\label{V-BS-kM}
V^{BS}_2=\left(\begin{array}{llll}
\eta_H t_{H,2}& i \eta_H r_{H,2} & 0 &0\\
i\eta_H r_{H,2} & \eta_H t_{H,2}& 0 & 0\\
0 & 0 & \eta_V t_{V,2}& i\eta_V r_{V,2}\\
0 & 0 & i\eta_V r_{V,2}& \eta_V t_{V,2}
\end{array}\right) ,
\end{equation}
which is, then,  used in the expression (\ref{MZ-general}) of $U^{real}_{\ba(\phi)}$.
Notice that, if the mirrors act differently on the two polarization channels, we can pass from the situation of equally lossy polarization described in Section \ref{secELPC} to the generic situation described in this section, and {\em viceversa}, depending on the value of $\eta_H$ and $\eta_V$.
\end{remark}

Let $\tilde e_{\ba,\bb}$ be defined as the difference between the detection probabilities \eqref{p-real-2} and the simplified ones \eqref{p-real-3}

\begin{multline}\label{etildeab}
\tilde e_{\ba,\bb}:=\Big| \frac{ \Tr[U^{real}_{\ba,\bb}\rho (U^{real}_{\ba,\bb})^\dag \, P^M_{x}\otimes P^P_{y}]  }{\Tr[U^{real}_{\ba,\bb}\rho (U^{real}_{\ba,\bb})^\dag ] }\\ -\Tr[\tilde U^{real}_{\ba,\bb}\rho (\tilde U^{real}_{\ba,\bb})^\dag \, P^M_{x}\otimes P^P_{y}] \Big|\, 
\end{multline}
and let $\tilde e$ be the supremum  over all possible choices of unit vectors $\ba,\bb \in \bR^3$ \begin{equation}\tilde e:=\sup_{\ba, \bb}\tilde e_{\ba,\bb}\label{etilde}\:.\end{equation}

The coefficient $\tilde e$ in (\ref{etilde}) depends also on the density matrix $\rho$. In order to take into account this, it is convenient to introduce the decomposition 
\begin{equation}\label{par-rho}
\rho = \alpha P_H\rho_HP_H + \beta P_V\rho_VP_V +P_V pP_H +P_Hp^\dagger P_V\:,
\end{equation}
where $P_H$ resp. $P_V$  are the two orthogonal projections operators onto the subspaces of $\cH_M\otimes \cH_P$ spanned by the vectors $\{ |0H\rangle, |1H\rangle\}$ resp. $\{ |0V\rangle, |1V\rangle \}$. The operators $\rho_H: \mathbb{C}^2_H \to \mathbb{C}^2_H$ and $\rho_V:\mathbb{C}^2_V \to \mathbb{C}^2_V$ are $2\times 2$ density matrices and $\alpha,\beta \geq 0$ with $\alpha+\beta =1$, whereas 
$p : \mathbb{C}^2_H \to \mathbb{C}^2_V$ is a linear operator. \\
By introducing the two constants $c_H$ and $c_V$ defined as 
\begin{align}
   c_H&=\sqrt{(t_{H,1}^2+r_{H,1}^2)(t_{H,2}^2+r_{H,2}^2)}  \nonumber \\
   c_V&=\sqrt{(t_{V,1}^2+r_{V,1}^2)(t_{V,2}^2+r_{V,2}^2)}\: ,\label{CHCV}
\end{align}
we have:
\begin{equation}\label{in-etilde}
    \tilde e \leq  \Big| \frac{\alpha\beta (c^2_H-c_V^2)}{\alpha c_H^2 +\beta c_V^2}\Big|+\left| \frac{\sqrt{\alpha \beta} (c_Hc_V - \alpha c_V^2 -\beta c_H^2)}{\alpha c_H^2 +\beta c_V^2}\right|
\end{equation}

The details of the derivation are postponed to appendix \ref{appendix-b}.

The values of the constants $\alpha, \beta$ can be estimated by taking into account the generation stage (I) (see fig. \ref{fig:Setup}). Indeed, the state $\rho$ of the photon entering the preparation stage (II)  is the result of the action of a collimator and a polarization filter selecting the vertical component, of a beam splitter, with transmission and reflection coefficients $t_{V,0}, r_{V,0}$, and a polarization rotator located along the reflected path and  converting the vertical polarization into the horizontal one.
If we  assume that the two mirrors in the generation stage  are lossy, but essentially identical to each other, the net effect of such non-ideal mirrors is just to change $t_{V,0}, r_{V,0}$ with a common factor $\eta \in (0,1)$.
According to this procedure, the generation stage produces a state $\rho$ of the form \eqref{par-rho} with coefficients $\alpha, \beta$ given by:   
\begin{equation}\label{alpha-beta}
\alpha =t_{V,0}^2/(t_{V,0}^2+r_{V,0}^2), \quad \beta =r_{V,0}^2/(t_{V,0}^2+r_{V,0}^2).
\end{equation}

Taking into account the analysis above and the one in the previous section, the difference between the detection probabilities, 
$P^{real}(x,y|\rho,\ba,\bb)$,
and those associated to product observables can be estimated by combining the found bounds as
\begin{multline}
    |P^{real}(x,y|\rho,\ba,\bb)-P^{ideal}(x,y|\rho,\ba,\bb)|\\
    \leq 2 \sqrt e +e +\tilde e \label{estimatePP}
\end{multline}
with $e$ given by \eqref{error-1} and $\tilde e$ bounded  by \eqref{in-etilde}. 
Analogously, the difference between the CHSH parameter \eqref{S-par} associated to the detection probabilities \eqref{p-real-2} 
and an ideal one associated to measurements of product observables on the same state $\rho$ can be estimated as 
\begin{multline}\label{Ris-1}
|S^{real}-S^{ideal}|\leq 4\sqrt 2(\|R_1\|+\|R_2\| ) \\ +2(\|R_1\|^2+\|R_2\|^2+\|R_1\|\|R_2\|) +16\tilde e
\end{multline}
where $\|R_1\|$ and $\|R_2\|$ are given by \eqref{R1R2norm}.\\

Sharper bounds can be obtained numerically as shown in \cite{Leone21}.

\begin{remark}
The bounds (\ref{estimatePP}) and (\ref{Ris-1}) have a precise  operative meaning.
The experimental setting represented  in Fig. \ref{fig:Setup} should ideally allow the  measurements of factorized observables  $\ba \cdot \boldsymbol{\sigma} \otimes \bb\cdot  \boldsymbol{\sigma}$, where $\ba$ and $\bb$ are initially  chosen in the Bloch sphere. 
Furthermore, from this  ideal perspective, the state of the photon  exiting the generation stage in Fig. \ref{fig:Setup} should be the Bell state $|\Psi_+\rangle=\frac{1}{\sqrt{2}}(|0H\rangle + i|1V\rangle)$.
However, the presence of   non-idealities in the optical components inevitably modifies the factorized form of the  actually measured observables on the one hand as well as the  state of the photon on the other hand.
 We can nevertheless extract from the experimental data the values of measurement of generic  factorized observables $\ba' \cdot \boldsymbol{\sigma} \otimes \bb \cdot \boldsymbol{\sigma}$ up to a certain error depending on the given  technical specifications of the components of the setup. As discussed in Remark \ref{remaa},  we can in particular  choose  $\ba'= \ba'(\phi,u_0,v_0)\equiv \ba_0$ in such a way that the error attains its minimal value. In this way the experimental data can be interpreted  as a measurement of  $\ba_0 \cdot \boldsymbol{\sigma} \otimes \bb \cdot \boldsymbol{\sigma}$, affected by the said minimal error, which is estimated in 
(\ref{estimatePP}) where $P^{ideal}$ is ascribed to $\ba_0\cdot\boldsymbol{\sigma} \otimes \bb\cdot \boldsymbol{\sigma}$. Referring to a set of four similar observables, the same argument leads to the estimate (\ref{Ris-1}) for the value of the CHSH parameter $S$.
\end{remark}

\section{Semi-device-independent entropy certification of a QRNG based on Bell inequality violation by SPE states}\label{sezQRNG}

Random numbers are a fundamental resource in several practical applications, ranging from Monte Carlo simulations to cryptography. In the latter case, the unpredictability of the sequence of random bits is an important issue, since it affects the security of the associated criptographic protocols. For this reason, it is important to certify randomness, i.e. to prove  that the random numbers are uniformly distributed, uncorrelated and unpredictable. The first two features can be rather easily checked by running suitable  statistical tests for the distribution of the  output string of random bits. On the other hand, the proof of unpredictability is a rather challenging problem that cannot be tackled  by properly tailored statistical tests. In principle a QRNG, whose entropy source is a quantum process, is more secure than a generic true random number generator since quantum physics is intrinsically probabilistic; most quantum phenomena are unpredictable and the theory allows to compute only the statistical distribution of the possible outcomes. In addition, from a practical point of view the detailed modelling of the underlying entropy source as well as the estimate of the min-entropy of the output string is robust and independent of additional classical noise sources \cite{frauchiger2013true}. In this framework, a rather challenging class  of QRNG recently proposed consists in  the so-called {\em device independent} QRNG, which in principle allow a certification  of the quality and the security of the random numbers they produce {\em independently of any detailed model of the devices}. The main figure of merit characterizing the randomness as well as the security of the output string in the criptographic applications  is the {\em guessing probability} $p_g$, defined as the probability of the most probable digit and the corresponding {\em min-entropy } $H_\infty$, given by $H_\infty=-\log_2p_g$ (see \cite{konig2009operational} for an operational meaning of this quantity). Hence, when a random number generator is used, it is important to provide a  certification that the output string contains a certain amount of min-entropy, in such way that the  application of a randomness extractor allows to obtain a sequence of almost-uniform random bits \cite{nisan1999extracting}.

Let's come back to  the framework described in section \ref{sez2}, where measurements of two commuting observables $\ba \cdot \boldsymbol{\sigma} \otimes I$ and $I\otimes \bb\cdot \boldsymbol{\sigma}$ are performed on a quantum system with two independent degrees of freedom and 
associated Hilbert space  $\cH_A\otimes\cH_B$ with $\cH_A=\cH_B=\bC^2$. 
 In the case where the state $\rho$ of the system  is a  pure state, i.e. $\rho =|\psi \rangle \langle \psi |$, with $\psi $ unit vector in $\cH_A\otimes \cH_B$, the amount of ``quantum randomness" contained in the measurement outcomes is quantified by the guessing probability $G(\psi, \ba, \bb)$ defined as  \begin{equation}
 G(\psi, \ba, \bb):=\max_{x,y }P(x,y|\psi,\ba,\bb) 
  \end{equation}
where $P(x,y|\psi,\ba,\bb) $ are given by \eqref{Q-dis}. 
An upper bound for $G(\psi, \ba, \bb)$ lower than $1$ certificates quantum randomness in the distribution $P$.
In the general case of a mixed state $\rho $, the {\em quantum guessing probability} is defined as 
   \begin{equation}\label{quantum-guessing}
 G(\rho, \ba, \bb)=\sup_{\{(\mu(\lambda),\psi_\lambda)_{\lambda\in\Lambda}\}}\int_\Lambda   G(\psi_\lambda, \ba, \bb) d\mu(\lambda)\:,\end{equation}
  where the sup is taken over all possible  decomposition   $\{(\mu(\lambda), \psi_\lambda)_{\lambda \in \Lambda}\}$  of $\rho$ into a convex superposition of pure states \footnote{identity \eqref{mixture} is meant in {\em weak sense}, where $\mu$ is any probability measure over a parameter  set $\Lambda$ and the vectors $\psi_\lambda$ are not mutually orthogonal in general}:
  \begin{equation}\label{mixture}\rho=\int_\Lambda  |\psi _\lambda\rangle \langle  \psi _\lambda | d\mu(\lambda)\:.\end{equation} 
  The idea behind  expression \eqref{quantum-guessing} is to separate  the quantum randomness of every pure state
$\psi_\lambda$  from its  {\em classical weight}  $\mu(\lambda)$, considering the exact value of the parameter $\lambda$ as a piece of information in principle accessible to a potential eavesdropper. As the notation itself says, the quantum guessing probability \eqref{quantum-guessing} has an explicit dependence on the quantum state  $\rho$ and on the couple of measured observables 
$\ba \cdot \boldsymbol{\sigma} \otimes I$ and $I\otimes \bb\cdot \boldsymbol{\sigma}$. A further step towards a more robust certification of the quantum randomness leads to the  {\em realization-independent quantum  guessing probability}  $G(P_{\rho,\ba,\bb})$ associated to a given distribution $$P_{\rho,\ba,\bb} :=\{P(x,y|\rho,\ba,\bb) \:|\: x,y\in \{-1,1\}\}$$ of the form \eqref{Q-dis}, which is defined by\begin{equation}
G(P_{\rho,\ba,\bb}):=\sup_{\{\tilde \rho,\tilde \ba, \tilde \bb \}}G(\tilde \rho, \tilde \ba, \tilde \bb), \label{real-Gp}
\end{equation}
where  the supremum is evaluated over 
all possible triples  $\{\tilde \rho,\tilde \ba, \tilde \bb\}$ of states $\tilde \rho$ and local observables $\tilde \ba \cdot \boldsymbol{\sigma} \otimes I$ and $I\otimes \tilde \bb\cdot \boldsymbol{\sigma}$ compatible with the  distribution $P_{\rho,\ba,\bb}$, i.e.: 
$$P(x,y|\rho,\ba,\bb)=Tr(\tilde{\rho}P_x^{\tilde{\ba}_i}\otimes P_y^{\tilde{\bb}_j}) \quad\mbox{for all $x,y\in \{-1,1\}$.}$$
This quantity actually provides a quantification of the amount of secure quantum randomness present in the distribution $P_{\rho,\ba,\bb}$ independently of any accessible (classical) side information and any particular description of the system. In fact $G(P_{\rho,\ba,\bb})$ depends only on the observed distribution of outcomes and gives a rather conservative bound on the probability of the most probable outcome in any (even the worst-case) scenario  under the only assumption that the quantum distribution $P$ is obtained from the measurement of observables in product form.
Correspondingly, the min-entropy $H_\infty :=-\log _2 G(P_{\rho,\ba,\bb})$ expresses this guessing probability in bits. In our specific  model with the above definition of $G(P_{\rho,\ba,\bb})$, $H_\infty$ varies in $[0,\alpha]$ where $\alpha$ is about $1.2$ (see the analysis in \cite{Pironio10} and the curve (a) in fig.2 therein, in particular). However, the upper bound (here $\alpha>1$) depends on the considered system and on the precise definition of $G(P_{\rho,\ba,\bb})$. In any cases, the more quantum randomness the analysed probability distribution contains, the more its entropy is positive and far from $0$. As a consequence, lower bounds for $H_\infty$ -- 
i.e., upper bounds for $G(P_{\rho,\ba,\bb})$ --
may be of crucial interest in applications.

To this regard, in \cite{acin2012randomness}, the authors considered  a Bell test (readapted to a quantum contextuality test in our case) with a measured CHSH parameter $S$ 
obtained 
by the measured probability distributions 
$P_{\rho,\ba,\bb}$
of four pairs  observables $\ba_i\cdot  \boldsymbol{\sigma} \otimes I$ and $I \otimes \bb_j \cdot \boldsymbol{\sigma}$, ($i,j=1,2$), 
on the state $\rho$. They proved that
the realization independent quantum guessing probability $G(P_{\rho,\ba,\bb})$ of each of the four measured distributions ($(\ba,\bb)\in \{(\ba_i,\bb_j)\:|\: i,j=1,2\}$) is bounded by 
\begin{equation}\label{pg-ideal}
   G(P_{\rho,\ba,\bb})\leq \frac{1}{2}+\frac{1}{2}\sqrt{2-\frac{S^2}{4}}\:.
\end{equation}
Note that the right-hand side gives rise to a meaningful (i.e.  $<1$) upper bound for $G(P_{\rho, \ba,\bb})$  only if  $2 < S \leq 2\sqrt{2}$, i.e. when $S$ stays between  the  CHSH classical  threshold  and the Tsirelson limit. 

In view of the discussion around the estimates (\ref{estimatePP}) and (\ref{Ris-1}),
in the case of a test of
quantum contextuality on SPE photons with realistic devices, inequality \eqref{pg-ideal} has to be modified in the following way 
\begin{equation}\label{pg-real}
    G(P^{real}_{\rho, \ba_i,\bb_j})\leq \frac{1}{2}+\frac{1}{2}\sqrt{2-\frac{(S^{real} -e_s ))^2}{4}}+e_p
\end{equation}
with $e_s= 4\sqrt 2(\|R_1\|+\|R_2\| )  +2(\|R_1\|^2+\|R_2\|^2+\|R_1\|\|R_2\|) +16\tilde e$ and $e_p= 2\sqrt e +e + \tilde e$, with $e$ given by \eqref{error-1}.
  This yields an equivalent lower bound for the associated min-entropy of 
  the considered measured probability distributions with
  the following form
  \begin{equation}\label{bound-Hmin} H_{\infty} \geq -\log_2\left( \frac{1}{2}+\frac{1}{2}\sqrt{2-\frac{(S^{real} -e_s )^2}{4}}+e_p\right)\end{equation}
for the min entropy of the probability distribution computed out of the raw data of (nominal) measurements of $\ba_i\cdot \boldsymbol{\sigma} \otimes \bb_j \cdot \boldsymbol{\sigma}$ with the apparatus of Fig. \ref{fig:Setup}, one of the four choices to obtain $S^{real}$ to insert in right-hand side.
These estimates are {\em semi-device independent} as, e.g.,  the right-hand side  depends only on the technical features of beam splitters and mirrors of the preparation stage (see Fig. \ref{fig:Setup}) embodied in $e_s$ and $e_p$.
The beam splitters  are here considered as quite realistic: they can  be different from each other,  lossy,  and  acting differently on states with different polarization, though they are assumed not to change the polarization of the photons they handle. 
The estimate \eqref{bound-Hmin} is robust under classical side information.

 In the practical implementation  of the protocol  described in \cite{Leone21}, the estimate of $S^{real}$ from the experimental data requires the fair sampling assumption.
In \cite{acin2012randomness}, the measured probabilities $P_{\rho,\ba,\bb}$ is evaluated by randomly choosing the value of $(\ba,\bb)$ for each round of the experiment. This is possible also within our semi-device independent protocol, but is not strictly necessary. A predetermined sequence of $(\ba,\bb)$ can be used as long as this knowledge cannot be exploited by an adversary to maliciously modify the outcomes of the experiment \cite{Pironio13}.

\section{SPE with realistic detectors}\label{sezMarkov}

In this section, we  analyse quantitatively realistic detectors and provide a technique for the construction of confidence intervals for the quantum probabilities \eqref{p-real-2} taking into account a finite number of experimental data. In particular, we develop a Markov model for the memory effects present in the sequence of measurement outcomes due to detector non idealities such as dead time, afterpulsing and dark counts rate (DCR). Finally, we propose an unbiased estimator for the quantum transition probabilities out of the collection of experimental data. This analysis plays an important role when the features of the  light source do not allow a strict control of photon arrival times, for example when an attenuated classical source is used as in \cite{Pasini20}. As anticipated in the introduction, the following analysis relies upon the fair sampling assumption. Specifically, in the application to the semi device independent QRNG protocol described in section \ref{sez4}, the provider of the detectors is assumed to be trusted.
 
 As explained in  \cite{Pasini20},  in an experiment of Bell inequality violation by SPE photons it is possible to associate to the  photons of the incoming beam a sequence of independent identically distributed random variables $\{\xi_n\}_{n\geq 1}$ with four possible outcomes $i=1,2,3,4$ associated to the four final channels  of the measuring apparatus, i.e. the four detectors, and the corresponding probabilities $p_i$.
We assume that the parameters characterizing the state preparation and the measurement stage are stable during the acquisition time, in such a way that the process $\{\xi_n\}_{n\geq 1}$ is stationary.

 In the ideal case, i.e. if we neglect afterpulsing and dead time and we assume a detection efficiency equal to  100\%, the sequence of measurement outcomes allows to estimate the theoretical probabilities $p_i$ in term of the corresponding empirical frequencies $N_i/N$, where $N_i$, $i=1,2,3,4$ is the number of counts on the $i$-th detector and  $N=\sum_iN_i$ is the total number of detected photons.  However, if we take into account the presence of detector non-idealities, such as dead time and afterpulsing, then we can no longer assume  the independence of the sequence of measurement outcomes and memory effects arise. Indeed, when a photon is detected,
the SPAD remains blind for a time interval of length $T_d$
where it is unable to detect additional photons that may reach
its sensitive volume.  On the other hand, there is a non-negligible probability of afterpulsing, i.e., of a subsequent readout caused by a secondary event produced in the SPAD instead of a further incoming photon.

In the following, we shall assume that the four detectors have equal efficiency $\eta \in (0,1]$ and shall take it into account by replacing the intensity $\lambda $ of the photon beam with an effective intensity $\lambda_{e }=\eta \lambda$. The probability of afterpulsing will be denoted by $p_a$ and assumed to be of  order $10^{-2}$ or less. The symbol $T_d$  will denote the detector dead time, while  $N(T_d)$  will be  the number of photons reaching the detectors during the dead time.  Let us call $\bP(N(T_d) \geq n)$ the probability that a number of photons greater or equal to $n$ reaches the detector  photons during the dead time. We shall assume  that $\bP(N(T_d)=1)=\epsilon $, with $\epsilon$  of  order $10^{-2}$ or less, and that $\bP(N(T_d)>1)=o(\epsilon)$. This condition is fulfilled, e.g., if the photon source is an attenuated laser, yielding a Poissonian distribution of the arrival times. In this case we have $\bP(N(T_d)=1)=\lae T_d e^{-\lae T_d }\sim \lae T_d$ and  $\bP(N(T_d)>1)=1- e^{-\lae T_d }-\lae T_d e^{-\lae T_d }$. If the expected value $\lae T_d$ of $N(T_d)$ is of order $10^{-2}$ we have $\bP(N(T_d)=1)\sim \lae T_d$ and  $\bP(N(T_d)>1)\sim (\lae T_d)^2/2=o(\lae T_d)$. 
Eventually, we shall assume that the time of afterpulsing $T_a$ is of the order of the dead time, i.e. $T_a\sim C\dt$ with $C=O(1)$ and correspondingly $\bP(N(T_a)=1)=O(\epsilon) $ and $\bP(N(T_a)>1)=o(\epsilon) $ .\\
Under these approximations it is possible to develop a Markov model for the correlations among  the subsequent readouts of the detectors  caused by non-idealities.\\
Let $\{\eta_n\}_{n\geq 1}$ be the sequence of random variables with 4 possible outcomes $i=1,2,3,4$ associated to the subsequent readouts of the four detectors.  Each realization of the sequence $\{\eta_n\}_{n\geq 1}$ actually gives the temporal sequence of outcomes of the measurements. In other words, with the statement $\eta_n=i$ we mean that the $n$-th data is collected by the $i$-th detector. 
In the ideal case, if dead time, DCR and afterpulsing are neglected, the two sequences $ \{\xi_n\}_{n\geq 1}$ and $ \{\eta_n\}_{n\geq 1}$ will have the same distribution. In the realistic case, the distribution of $ \{\eta_n\}_{n\geq 1}$ is actually affected by the non-idealities of the measuring apparatus in the way we are going to describe.
Let us consider first the case where DCR gives a negligible contribution and focus on 
 the first detected  photon. Since in this case  neither of the four detectors is in dead time nor in afterpulsing caused by previous detections,  the first variable $\eta_1$ has the same distribution of $\xi_1$, i.e. 
$\bP(\eta_1=i)=p_i$, $i=1,...,4$.
Let us consider now the second detection, whose statistics is described by the random variable $\eta_2$.  The correlations between $\eta_1$ and $\eta _2$ are described by the set of conditional probabilities $\bP(\eta_2=j|\eta_1=i)$, with $i, j=1,...,4$.   Taking into account afterpulsing, whose occurrence is denoted with the symbol $AFP$ (while $AFP^c$ denotes the complementary event), we have 
\begin{multline*}
    \bP(\eta_2=j|\eta_1=i)
=\bP(\eta_2=j|\eta_1=i\cap AFP)p_a\\+\bP(\eta_2=j |\eta_1=i \cap  AFP^c)(1-p_a).
\end{multline*} 
We have implicitly  assumed that all detectors have the same probability of afterpulsing 
$\bP(AFP|\eta_1=i)=p_a,$  $ i =1,\ldots,4$.
Concerning the first term, denoting by $\tau$ the interarrival time between the first and the second photon, we have:
\begin{multline*}
    \bP(\eta_2=j|\eta_1=i\cap AFP)\\
=\bP(\eta_2=j|\eta_1=i\cap AFP\cap \tau<T_a)\bP( \tau<T_a|\eta_1=i\cap AFP)\\ +\bP(\eta_2=j|\eta_1=i\cap AFP\cap \tau>T_a)\bP( \tau>T_a|\eta_1=i\cap AFP)\\
=\bP(\eta_2=j|\eta_1=i\cap AFP\cap \tau<T_a)\bP(N(T_a)>0)\\ +\delta_{ij}\bP(N(T_a)=0)
\end{multline*}

Now, observing  that this term has to be multiplied by $p_a\sim 10^{-2}$, in the case where  $\bP(N(T_a)>0)=O(\epsilon)$, we can neglect the first term obtaining:
$$ \bP(\eta_2=j|\eta_1=i\cap AFP)\sim \delta_{ij}\bP(N(T_a)=0).$$
Let us consider now the probability $\bP(\eta_2=j |\eta_1=i \cap  AFP^c)$ that, given the  result of the first measurement is $i$ and no afterpulsing occurs, the result of the second measurement is $j$. Denoting with $N(T_d)$ the number of photons reaching the detectors during the dead time, we have
\begin{align*} 
&\bP(\eta_2=j |\eta_1=i \cap  AFP^c)\\
&=\sum_{k=0}\bP(\eta_2=j \cap N(T_d)=k|\eta_1=i \cap  AFP^c)\\
&=\sum_{k=0}\bP(\eta_2=j |\eta_1=i \cap  AFP^c \cap N(T_d)=k)\bP(N(T_d)=k)\\
&=\bP(\eta_2=j |\eta_1=i \cap  AFP^c \cap N(T_d)=0)\bP(N(T_d)=0)\\
 & +\bP(\eta_2=j |\eta_1=i \cap  AFP^c \cap N(T_d)=1)\bP(N(T_d)=1)\\
  & \qquad \qquad \qquad \qquad\qquad \qquad \qquad \qquad \qquad +o(\epsilon)\\
 &=p_j (1-\epsilon +o(\epsilon))+ q_{ij}\epsilon+o(\epsilon)
\end{align*}
where, if $i=j$,
$$q_{ij}=\bP(\eta_2=j |\eta_1=i \cap  AFP^c \cap N(T_d)=1)=p_j^2,$$
while, if $i\neq j$,
$$q_{ij}=\bP(\eta_2=j |\eta_1=i \cap  AFP^c \cap N(T_d)=1)=p_j +p_ip_j$$
Under the assumption that $\epsilon$ is so small than we can neglect all the terms of order $o(\epsilon)$, the conditional probabilities satisfy the Markov property
\begin{multline}\label{Markov}\bP(\eta_{n+1}=i_{n+1}|\eta_1=i_1, \ldots, \eta_{n}=i_n )\\=\bP(\eta_{n+1}=i_{n+1}|\eta_{n}=i_n )=\bP(\eta_{2}=i_{n+1}|\eta_{1}=i_n ),\end{multline}
since, in fact, the left hand side of \eqref{Markov} is equal to the right hand side plus additional terms which account for  the cases where $N(T_d)\geq 2$ and these have a negligible probability. In summary,  according to  our approximations, the sequence of random variables $\{\eta_n\}_{n\geq 1}$ is a stationary Markov chain with transition probabilities $\bP(\eta_{n+1}=j|\eta_{n}=i )=P_{ij}$ given (up to term of order $o(\epsilon)$ by:
\begin{equation}\nonumber P_{ij}=p_a\delta_{ij}+(1-p_a )\left((1-\epsilon) p_j+\epsilon q_{ij}\right)\end{equation}
Equivalently, the stochastic matrix  $P$ is equal to  \begin{equation}\nonumber P=p_aI_{4 \times 4}+(1-p_a )\left((1-\epsilon) \tilde P+\epsilon Q\right),\end{equation} with 
\[ \tilde P=\left(\begin{array}{llll}
p_1 & p_2 & p_3 & p_4\\
p_1 & p_2 & p_3 & p_4\\
p_1 & p_2 & p_3 & p_4\\
p_1 & p_2 & p_3 & p_4
\end{array}\right)\] \[ Q=\left(\begin{array}{llll}
p_1 ^2& p_2 (1+p_1) & p_3 (1+p_1)& p_4(1+p_1)\\
p_1 (1+p_2) & p_2^2 & p_3 (1+p_2) & p_4(1+p_2) \\
p_1 (1+p_3) & p_2 (1+p_3) & p_3^2 & p_4 (1+p_3) \\
p_1 (1+p_4) & p_2 (1+p_4)  & p_3  (1+p_4)  & p_4^2
\end{array}\right)\]
If  $p_i>0$ for all $i=1,2,3,4$, then the Markov chain is  irreducible and  by the ergodic theorem the empirical frequencies converge to the unique invariant distribution $(f_i)_{i=1,\ldots 4}$. More precisely, by denoting $N^{i}_n:=\sum _{k=1}^n 1_{\eta_k=i}$, we have 
$$\bP\left( \lim_{n\to \infty }\frac{N^i_n}{n}- f_i=0\right)=1.$$
where $(f_1,f_2,f_3,f_4)$ can be computed as the left eigenvector of the matrix $P$ with eigenvalue 1:
$$f_i=\frac{p_i}{1+\epsilon p_i}\left(\sum _{j=1}^4 \frac{p_j}{1+\epsilon p_j}\right)^{-1}
\sim p_i+\epsilon\,  p_i(\sum_jp_j^2-p_i)$$
The latter relation can be easily inverted up to terms of order $o(\epsilon)$ yielding:
\begin{equation}\nonumber p_i\sim f_i\left(1+\epsilon\left(f_i+\sum _{j=1}^4f_j^2\right)\right)\end{equation}
This formula provides a rough estimator of the theoretical probabilities $p_i$ in terms of the empirical frequencies $f_i$, $i=1,...,4$.\\
More precise and unbiased estimators for the parameters $p_i$, $i=1,...,4$, as well as the corresponding confidence intervals can be obtained via the maximum likelihood principle. Given a realization of the Markov chain described above, i.e. a sequence of outcomes $\{x_i\}_{i=1,\ldots, n}$, with $x_i=1,2,3,4$, its probability is given by 
$$p_{x_1}\prod_{i=1}^{n-1}P_{x_ix_{i+1}},$$
and the corresponding log-likelihood is equal to 
\begin{align}\label{log-lik}l(P)&:=\log(p_{x_1}\prod_{i=1}^{n-1}P_{x_ix_{i+1}})\\ &=\log (p_{x_1})+\sum_{i,j=1,2,3,4}N_{ij}\log(P_{ij}),\end{align}
where $N_{ij}$ is the number of transitions from $i$ to $j$. 
The estimated values of the parameters $p_i$ as well as the corresponding confidence interval can be obtained by maximization of \eqref{log-lik} under the four constraints $\sum_jP_{ij}=1$, $i=1,...,4$ (see \cite{billingsley1961statistical,ammicht1982maximum} for the underlying theory). A practical implementation of this technique is presented in \cite{Leone21}, where the the statistical programming language R has been used.\\
Eventually, if DCR is not negligible then the model described above can be easily updated by replacing the distribution of the initial random variables $\xi_n$
$$\bP(\xi_n=i)=p_i, \qquad i=1,2,3,4,$$
with the corrected values 
\begin{equation}\nonumber \bP(\xi_n=i)=\tilde p_i=(1-p_{DCR}) p_i+\frac{p_{DCR}}{4}, \qquad i=1,2,3,4,\end{equation}
where $p_{DCR}$ is the total fraction of detected photons due to dark counts.

For an application of this technique to an experiment of CHSH violation by SPE photons coming from an attenuated laser source see \cite{Leone21}.

\section{Conclusions}\label{conclusions}

This work shows how the entropy of a realistic QRNG based on momentum-polarization single-photon entangled states can be certified despite the non-idealities of the employed optical and electronic devices. Our analysis starts by taking into account the polarization-dependent responses of beam splitters and mirrors to estimate, first, the real conditional probabilities $P^{real}_{\rho,\ba,\bb}$ due to the optical setup, and, from that, the real CHSH correlation function $S^{real}$. Then, an upper bound $e_P$ is carefully evaluated between the real conditional probabilities and the ideal ones. Consequently, an upper bound $e_S$ is estimated for the distance $|S^{real}-S^{ideal}|$ between the real and the ideal CHSH functions used for proving the entanglement. According to device independent protocols where the violation of the CHSH inequality ensures a level of min-entropy, $S^{real}$ together with $e_P$ and $e_S$ concur to finally define a modified expression for the certified lower value  of the min-entropy. The resulting certified QRNG falls into the class of the \emph{semi-device independent} ones, as a modeling is needed due to some feature of the preparation stage (beam splitters and mirrors optical responses), but not needed for others (momentum and polarization angles). Moreover, the fact that single photons are measured with realistic detectors having non-unitary efficiency and affected by dead time, afterpulsing and dark counts, is here considered by means of a Markovian model used to take memory effects into account by correcting for the measured probabilities and allowing for the construction of the corresponding confidence intervals tackling the issue of finite statistics.
The model implicitly relies on the fair sampling assumption and on the hypothesis that  the preparation and measurement parameters of the system  are stable during the acquisition time.  Finally, the present QRNG protocol is robust under classical side information (see \cite{Pironio13}). We plan to generalize it to the case of quantum side information  in a future paper along the lines of, e.g.,  \cite{arnon2018practical,zhang2020efficient}.

Our analysis demonstrates that single-photon entanglement is not simply interesting from a fundamental point of view \cite{azzini2020single}, but it can be a practical resource in quantum information, thanks in particular to the fact that attenuated classical light sources can be used \cite{Pasini20}. An example of application of entropy certification of a QRNG based on entangled single photons from a weak laser beam is demonstrated by the experiment reported in \cite{Leone21}. Nevertheless, our analysis of optical non-idealities can also be applied to experimental tests of
quantum contextuality of single-particle entangled states generated from heralded single photons \cite{Michler2000,Gadway09,Chen10,Karimi10}, where the photon spin or polarization is necessarily one of the two degrees of freedom involved. Indeed, in these cases as well, beam splitters and mirrors - the optical source of an unwanted coupling between the different degrees of freedom - are needed to build a Mach-Zehnder interferometer serving as a gate for the other employed qubit, be it linear \cite{Michler2000,Gadway09} or angular \cite{Chen10,Karimi10} momentum. In this sense, our results could be exploited to enable the certification of the entropy of a QRNG based on single-photon entanglement involving degrees of freedom others than momentum and polarization. And even more generally, they could potentially be of interest to tests of
quantum contextuality of single-particle entanglement exploiting particles other than photons, e.g. neutrons \cite{GeppertHasegawa2014,Shen20} or atoms \cite{Jeske}.

%
%

\begin{acknowledgements}
We acknowledge helpful discussions with C. Agostinelli on the code implementation of the Markovian model. This project has received funding from the European Union’s Horizon 2020 research and innovation
programme under grant agreement No 820405. NL was supported by a fellowship of Q@TN within the PAT(AI) grant.

\end{acknowledgements}

\appendix*
\section{Proof of equation \eqref{equalitymaxmin}}\label{sect-proof-equalitymaxmin}
Let us compute the minimum Hilbert-Schmidt distance between the operator  $\tilde U_\ba^{real}$ given by \eqref{U-real-block} (denoted below by $\tilde{U}^{real}$ for shortness) and a  unitary operator of the form $A\otimes I$, with $A=e^{i\zeta}e^{i\alpha \, \hat m\cdot \boldsymbol{\sigma}}$.
By explicit computation we have:
\begin{multline}\|\tilde U^{real}-(A\otimes I)\|_{HS}^2=\Tr[(|\tilde U^{real}-(A\otimes I))(|\tilde U^{real}-(A\otimes I)^\dag)]\\
=8-4\cos \zeta(\cos\alpha (\cos \theta_H+\cos\theta_V)\\ 
+\sin \alpha\, \hat m \cdot (\sin \theta_H \, \hat n_h + \sin \theta_V \, \hat n_V))
\end{multline}
Clearly, the minimum of the distance $\|\tilde U^{real}-(A\otimes I)\|_2$ is attained for those values of $\zeta, \alpha, \hat m$ that maximize the function $\cos (\zeta)\,  f(\alpha, \hat m)$, with $f(\alpha , \hat m):=\cos\alpha (\cos \theta+\cos\theta')+\sin \alpha\, \hat m \cdot (\sin \theta \, \hat n + \sin \theta' \, \hat n')$. Since $\cos (\zeta)\in [-1,+1]$ the problem is reduced to maximize the absolute value of $f(\alpha , \hat m)$, for $\alpha \in [0,2\pi]$ and $\hat m\in \bR^3$, $\|\hat m\|=1$. By direct computation we obtain:
\begin{multline*}
\max_{\alpha \in [0,2\pi], \|\hat m\|=1}\left|f(\alpha , \hat m)\right|
=(2+2\cos^2\phi/2\cos(\alpha +\beta -\alpha '-\beta')\\ +2\sin^2\phi/2 \cos(\alpha -\beta -\alpha '+\beta'))^{1/2}
\end{multline*}
Hence, by direct computation 
\begin{multline*}
\max_{\phi/2\in [0,2\pi]}\min_{\alpha \in [0,2\pi], \|\hat m\|=1}\|\tilde U^{real}-(A\otimes I)\|_2^2\\
=8\Big(1-\min\big\{\left|\cos\left(\frac{\alpha_1^H +\alpha_2^H -\alpha_1^V-\alpha_2^V}{2}\right)\right|,\\  \left|\cos\left(\frac{\alpha_1^H -\alpha_2^H -\alpha_1^V+\alpha_2^V}{2}\right)\right|\big\}\Big)
\end{multline*}
and this final result coincides with \eqref{error-1}.

More generally, we can 
 consider  the  case where the unitary operator $A\otimes I$, with $A= e^{i\zeta}U_M$,  $\zeta \in [0,2\pi]$ and $U_M\in SU(2)$,  is replaced by  the general unitary operator in the product form $A\otimes U_P$,  $U_P\in SU(2)$. We shall use the notation $U_M=e^{i\alpha \, \hat m\cdot \boldsymbol{\sigma}}$ and $U_P=e^{i\beta \, \hat k\cdot \boldsymbol{\sigma}}$, $\alpha, \beta \in [0,2\pi]$, $\hat m, \hat k\in \bR^3$, $\|m\|=\|k\|=1$. In particular, if $U_P$ has the form
\begin{align*}U_P&=\left(\begin{array}{ll}
u_{HH}& u_{HV} \\
u_{VH} & u_{VV}
\end{array}\right) \\ &=\left(\begin{array}{ll}
\cos\beta +i\sin \beta\, k_z& \sin\beta (i k_x+k_y) \\
 \sin\beta (i k_x-k_y)  & \cos\beta -i\sin \beta\, k_z
\end{array}\right)\end{align*}
the matrix associated to the  tensor product $A\otimes U_P$ assumes the following block form 
$$\left(\begin{array}{ll}
u_{HH}A& u_{HV}A \\
u_{VH} A& u_{VV}A
\end{array}\right) $$

Analogously as before, we look for the minimum of the Hilbert-Schmidt distance between  the opearator $\tilde U^{real}=\left(\begin{array}{ll}
U(\theta, \hat n) & 0 \\
0 & U(\theta', \hat n')
\end{array}\right) $ and $A\otimes U_P$. By explicit computation, using the identity
\begin{multline*}|u_{HH}|^2+|u_{VV}|^2+|u_{VH}|^2+|u_{HV}|^2\\ =2\cos^2\beta+2\sin^2 \beta \, k_z^2+2\sin^2\beta\, (k_x^2+k_y^2)=2\end{multline*} we get
\begin{multline}\nonumber
\|\tilde U^{real}-(A\otimes U_P)\|_2^2\\=\Tr[(|\tilde U^{real}-(A\otimes U_P))(|\tilde U^{real}-(A\otimes U_P)^\dag)]\\
=\Tr[4I_{2\times 2}-(u_{HH} U(\theta, \hat n)^\dag A+h.c.)-(u_{VV}U(\theta', \hat n')^\dag A+h.c)].
\end{multline}
Hence, we have to find the values of $\zeta, \alpha , \beta \in [0,2\pi]$ and $ \hat m, \hat k\in \bR^3$, $\|\hat m\|=\|\hat k\|=1$, maximizing the function 
 \begin{multline}\nonumber
 g(\zeta, \alpha , \beta,  \hat m, \hat k):=
 \Tr[(u_{HH} U(\theta_H, \hat n_H)^\dag A+h.c.)\\ +(u_{VV}U(\theta_V, \hat n_V)^\dag A+h.c\\
 = \cos \zeta\cos \beta\big(\cos\alpha(\cos\theta_H +\cos \theta_V)+\sin\alpha \,\hat m\cdot(\sin\theta_H \, \hat n_H\\ +\sin\theta _V\, \hat n_V)\big)\\ -\sin \zeta \sin\beta k_z\big(\cos\alpha(\cos\theta_H -\cos \theta_V)\\  +\sin\alpha \,\hat m\cdot(\sin\theta_H \, \hat n_H -\sin\theta _V\, \hat n_V)\big)
 \end{multline} 

By explicit computation we get:
$$\max_{\zeta, \alpha , \beta,  \hat m, \hat k}g(\zeta, \alpha , \beta,  \hat m, \hat k)=\max_{\alpha , \hat m}|f(\alpha , \hat m)|$$ which yields again the same result in \eqref{error-1}.

\section{Proof of inequality \eqref{in-etilde}}\label{appendix-b}

It is convenient to introduce the following notation for later use.
Let $V_H, V_V$ and be the matrices defined as
\begin{align*}
    V_H&:=\left(\begin{array}{ll}
 t_{H,1}& i r_{H,1} \\
i r_{H,1}&  t_{H,1}
\end{array}\right) V(\phi)\left(\begin{array}{ll}
 t_{H,2}& i r_{H,2} \\
i r_{H,2} & t_{H,2} 
\end{array}\right), \\
V_V&:=\left(\begin{array}{ll}
 t_{V,1}& i r_{V,1}\\
i r_{V,1} &  t_{V,1}
\end{array}\right) V(\phi)\left(\begin{array}{ll}
 t_{V,2}& i r_{V,2}\\
i r_{V,2} &  t_{V,2}
\end{array}\right)
\end{align*}
with $V(\phi)$ given by \eqref{Vphi}.
Analogously, let  $U_H, U_V$ be the matrices defined as 
$$U_H:=\frac{V_H}{c_H}, \quad U_V:=\frac{V_V}{c_V}$$
where  $c_H$ and $c_V$ are  the two positive constant given by \eqref{CHCV}. We actually have:
\begin{align*}U_H&:=\left(\begin{array}{ll}
\tilde t_{H,1}& i\tilde r_{H,1} \\
i\tilde r_{H,1} & \tilde t_{H,1} 
\end{array}\right) V(\phi)\left(\begin{array}{ll}
\tilde t_{H,2}& i\tilde r_{H,2} \\
i\tilde r_{H,2} & \tilde t_{H,2}
\end{array}\right)\\
U_V&:=\left(\begin{array}{ll}
\tilde t_{V,1}& i\tilde r_{V,1} \\
i\tilde r_{V,1} & \tilde t_{V,1} 
\end{array}\right) V(\phi)\left(\begin{array}{ll}
\tilde t_{V,2}& i\tilde r_{V,2} \\
i\tilde r_{V,2} & \tilde t_{V,2}
\end{array}\right)
\end{align*}
In particular the operators $\tilde U^{real}_{\ba,\bb}=(I\otimes U_\bb)\otimes U^{real}_\ba$ and $U^{real}_{\ba,\bb}$, defined respectively in \eqref{MZ-general} and \eqref{utildereal-bis}, can be represented as
\begin{align}\tilde U^{real}_{\ba,\bb}&=(I\otimes U_\bb)(P_HU_HP_H+P_V U_VP_V),\nonumber \\  U^{real}_{\ba,\bb}&=(I\otimes U_\bb)(c_HP_HU_HP_H+c_VP_VU_VP_V) \label{notation-1}\end{align}
where $P_H$ resp. $P_V$  are the  projections operators on the subspaces of $\cH_M\otimes \cH_P$ spanned by the vectors $\{ |0H\rangle, |1H\rangle\}$ resp. $\{ |0V\rangle, |1V\rangle \}$.

By introducing the notation $P^{P,\bb}_{y}:=U_\bb^+P^P_{y}U_\bb$, and using \eqref{notation-1}, we get that the difference parameter $\tilde e_{\ba,\bb}$ defined in \eqref{etildeab} can be estimated as:
\begin{multline*}
\tilde e_{\ba,\bb}
=\Big| \Tr\Big[ \Big(\left(\frac{c^2_H}{D}-1\right)P_HU_HP_H\rho P_HU^\dagger_HP_H \\ +\left(\frac{c_Vc_H}{D}-1\right)\left(P_VU_VP_V
\rho P_HU^\dag_HP_H+h.c.\right) \\ +\left(\frac{c^2_V}{D}-1\right)P_VU_VP_V\rho P_VU^\dag_VP_V \Big)P^M_{x}\otimes P^{P,\bb}_{y}\Big]  \Big|
\end{multline*}
where $D=c_H^2\Tr[P_H\rho P_H]+c_V^2\Tr[P_V\rho P_V]$.

Let $\cR_\ba$ be the operator defined as
\begin{multline*}
\cR_\ba:= \Big(\left(\frac{c^2_H}{D}-1\right)P_HU_HP_H\rho P_HU^\dag_HP_H \\ +\left(\frac{c_Vc_H}{D}-1\right)\left(P_VU_VP_V
\rho P_HU^\dag_HP_H+h.c.\right) \\ +\left(\frac{c^2_V}{D}-1\right)P_VU_VP_V\rho P_VU^\dag_VP_V \Big)
\end{multline*}
in such a way that, e.g., $\tilde e_{\ba,\bb}\leq \|\cR_\ba\|$ for any choice of $\ba, \bb$.\\
By adopting the decomposition  \eqref{par-rho} for the density matrix $\rho$:
\begin{equation*}
\rho = \alpha P_H\rho_HP_H + \beta P_V\rho_VP_V +P_V pP_H +P_Hp^\dagger P_V\:,
\end{equation*}
where $\rho_H: \mathbb{C}^2_H \to \mathbb{C}^2_H$ and $\rho_V:\mathbb{C}^2_V \to \mathbb{C}^2_V$ are $2\times 2$ density matrices and $\alpha,\beta \geq 0$ with $\alpha+\beta =1$, whereas 
$p : \mathbb{C}^2_H \to \mathbb{C}^2_V$. 
The positivity of the density matrix  $\rho$ yields 
$$|\langle x, p y \rangle|^2 \leq    \langle x, \alpha \rho_H x \rangle \langle y, \beta \rho_V y\rangle\quad \forall x,y \in \mathbb{C}^2\equiv \mathbb{C}^2_H \equiv \mathbb{C}^2_V\:. $$
In particular $p$ vanishes when either $\alpha=0$ or $\beta=0$. Therefore, it is safe  to rename $p$ in the previous decomposition as $\alpha\beta p$.  The above decomposition now reads 
\begin{equation*}
\rho = \alpha P_H\rho_HP_H + \beta P_V\rho_VP_V +\sqrt{\alpha \beta}P_V pP_H +\sqrt{\alpha \beta}P_Hp^\dagger P_V\:,
\end{equation*}
where  $\alpha,\beta, \rho_H, \rho_V$ are as above and  $p : \mathbb{C}^2_H \to \mathbb{C}^2_V$ is an arbitrary operator satisfying
\begin{equation}|\langle x, p y \rangle|^2 \leq    \langle x,  \rho_H x \rangle \langle y,  \rho_V y\rangle\quad \forall x,y \in \mathbb{C}^2\equiv \mathbb{C}^2_H \equiv \mathbb{C}^2_V\:. \label{condp}\end{equation}
By exploiting this decomposition, the operator $\cR_\ba$ can be rephrased to
\begin{equation*}
\cR_\ba=g_1(\alpha,\beta) Q_1+g_2(\alpha,\beta) Q_2
\end{equation*}
where 
$$g_1(\alpha,\beta):= \frac{\sqrt{\alpha \beta} (c_Hc_V - \alpha c_V^2 -\beta c_H^2)}{\alpha c_H^2 +\beta c_V^2}\:,$$ $$  Q_1:= \left(\begin{array}{ll}
0&U_Hp^\dagger U_V^\dagger\\
U_V p U_H^\dagger & 0
\end{array}\right)\:,$$
and 
$$g_2(\alpha,\beta):= \frac{\alpha\beta (c^2_H-c_V^2)}{\alpha c_H^2 +\beta c_V^2}$$
$$Q_2=\left(\begin{array}{ll}
U_H \rho_HU^\dag _H&0\\
0 & -U_V \rho_VU^\dag_V
\end{array}\right)\:.$$
As a consequence
$$||g_1Q_1||^2 = g_1^2 ||Q^\dagger Q|| \:.$$
The last norm can be estimated  observing that 
$$Q_1^\dagger Q_1:= \left(\begin{array}{ll}
U_Hp^\dagger p U_V^\dagger&0\\
0&U_Vp p^\dagger U_H^\dagger
\end{array}\right)$$ hence  $$ ||Q_1^\dagger Q_1||= \max \{||U_Hp^\dagger p U_V^\dagger||, ||U_Vp p^\dagger U_H^\dagger||\}$$
$$=  \max \{||p^\dagger p ||, ||p p^\dagger||\}= \max\{||p||^2, ||p^\dagger||^2\}  = ||p||^2\:.$$
From (\ref{condp}), with $y\in \mathbb{C}^2$ with $||y||=1$
$$||py||^4 =|\langle py, p y \rangle|^2 \leq    \langle py,  \rho_H py \rangle \langle y,  \rho_V y\rangle= ||\sqrt{\rho_H}py||^2\: ||\sqrt{\rho_V}y||^2 $$
hence
$$||py||^4 \leq ||\sqrt{\rho_H}||^2||py||^2  ||\sqrt{\rho_V}y||^2,$$
so that
$$||p||^2 \leq ||\sqrt{\rho_H}||^2 ||\sqrt{\rho_V}||^2 = ||\rho_H||\:||\rho_V||\:,$$
and eventually 
$$||g_1(\alpha,\beta)Q_1||\leq |g_1(\alpha,\beta)| \sqrt{||\rho_H||\:||\rho_V||}$$
The operator $Q_2$ is in block form and its eigenvalues can be computed as the eigenvalues of the two blocks, which are  given by density matrices (with eigenvalues bounded by 1). This allows to conclude that for any choice of $\rho$ the operator norm of this term is bounded by 1. Analogously, a similar estimate can be obtained for the term
$| \sqrt{||\rho_H||\:||\rho_V||}$. 
This gives
\begin{equation}\label{dis-fin-0}\|\cR_\ba\|\leq |g_1(\alpha, \beta)| +|g_2(\alpha, \beta)|  \end{equation}
where now the constants $g_i(\alpha, \beta)$, $i=1,2$, do not depend explicitly on $\phi$, or equivalently on  $\ba$, and we eventually obtain \eqref{in-etilde}.
\bibliography{SPE-QRNG-theory}
\end{document}